\documentclass[aps,pre,twocolumn,superscriptaddress,showpacs,showkeys,amsmath,amssymb,footinbib,longbibliography]{revtex4-1}
\usepackage[T1]{fontenc}
\usepackage[utf8]{inputenc}
\usepackage{textcomp}
\usepackage{multirow}
\usepackage{float}
\usepackage{color}
\usepackage{bm}
\usepackage{amsmath}
\usepackage{amssymb}
\usepackage{graphicx}

\makeatletter

\usepackage{latexsym}
\usepackage[ngerman,english]{babel}

\newcommand\mathplus{+}

\renewcommand{\theequation}{\arabic{section}.\arabic{equation}}

\begin{document}

\title{Thermodynamics of the pyrochlore-lattice quantum Heisenberg antiferromagnet}

\author{Patrick M\"{u}ller}
\affiliation{Institut f\"{u}r Physik,
          Otto-von-Guericke-Universit\"{a}t Magdeburg,
          P.O. Box 4120, 39016 Magdeburg, Germany}

\author{Andre Lohmann}
\affiliation{Institut f\"{u}r Physik,
          Otto-von-Guericke-Universit\"{a}t Magdeburg,
          P.O. Box 4120, 39016 Magdeburg, Germany}

\author{Johannes Richter}
\affiliation{Institut f\"{u}r Physik,
          Otto-von-Guericke-Universit\"{a}t Magdeburg,
          P.O. Box 4120, 39016 Magdeburg, Germany}
\affiliation{Max-Planck-Institut f\"{u}r Physik komplexer Systeme, 
          N\"{o}thnitzer Stra\ss e 38, 01187 Dresden, Germany}       
    
\author{Oleg Derzhko}
\affiliation{Institute for Condensed Matter Physics,
          National Academy of Sciences of Ukraine,
          Svientsitskii Street 1, 79011 L'viv, Ukraine}
\affiliation{Max-Planck-Institut f\"{u}r Physik komplexer Systeme, 
          N\"{o}thnitzer Stra\ss e 38, 01187 Dresden, Germany}       
\affiliation{Department for Theoretical Physics,
          Ivan Franko National University of L'viv,
          Drahomanov Street 12, 79005 L'viv, Ukraine}
                    
\date{\today}

\begin{abstract}
We use the rotation-invariant Green's function method (RGM) and the high-temperature expansion (HTE) 
to study the thermodynamic properties of the Heisenberg antiferromagnet on the pyrochlore lattice. 
We discuss the excitation spectra as well as various thermodynamic quantities,
such as spin correlations, uniform susceptibility, specific heat and static and dynamical structure factors.
For the ground state we present RGM data for arbitrary spin quantum numbers $S$.
At finite temperatures we focus on the extreme quantum cases $S=1/2$ and $S=1$. 
We do not find indications for magnetic long-range order for any value of $S$.
We discuss the width of the pinch point in the static structure factor in dependence on temperature and spin quantum number.   
We compare our data with experimental results for the pyrochlore compound NaCaNi$_2$F$_7$ ($S=1$).
Thus, 
our results for the dynamical structure factor agree well with the experimentally observed features at 3 \ldots 8~meV for NaCaNi$_2$F$_7$.
We analyze the static structure factor ${S}_{\bf q}$ to find regions of maximal ${S}_{\bf q}$. 
The high-temperature series of the ${S}_{\bf q}$ provide a fingerprint of weak {\it order by disorder} selection of a collinear spin structure,  
where (classically) the total spin vanishes on each tetrahedron and neighboring tetrahedra are dephased by $\pi$.
\end{abstract}

\pacs{
75.10.-b, 
75.10.Jm  
}

\keywords{quantum Heisenberg antiferromagnet, 
pyrochlore lattice, 
rotation-invariant Green's function method, 
high-temperature expansion, 
structure factor}

\maketitle

\section{Introduction}
\label{sec1}

Geometrically frustrated magnetic materials are a subject of great interest nowadays.
Phenomena of geometric frustration may emerge 
if nearest-neighbor antiferromagnetic interactions occur in periodic lattices based on triangles as elementary objects of the lattice structure
since the spins within a triangular cell cannot be mutually antiparallel.
One of the most prominent spin model in the field of geometrically frustrated magnetism 
is the pyrochlore Heisenberg antiferromagnet (PHAF).
The pyrochlore lattice is a three-dimensional arrangement of corner-sharing tetrahedra, 
see Fig.~\ref{fig01}, below. 
There are several families of compounds in nature 
with magnetic atoms which reside on the pyrochlore-lattice sites and interact
with their neighbors through antiferromagnetic exchange interactions,
see, e.g., Refs.~\cite{Gardner_2010,Gingras_McClarty_2014,Rau2018}.
On the other hand,
this spin model presents a playground for the study of geometric frustration in three dimensions. 
It is highly nontrivial and is far from being fully understood. 
Even in the classical limit there is no magnetic order 
and the ground state is a classical spin liquid with algebraically decaying spin-spin correlations \cite{Isakov_2004,Henley_2010}.
For low spin quantum numbers $S$ the complexity of the model increases, since quantum fluctuations become important.
Thus, so far for the quantum model no accurate values for the ground-state energy are available. 
At finite temperatures, 
the interplay of quantum and thermal fluctuations makes a theoretical investigation even more challenging.
While for the classical PHAF several accurate numerical tools available 
(e.g., Monte Carlo and molecular dynamics),  
such straightforward numerical tools do not work for the quantum PHAF.

Let us mention here two other models, which will be used below to compare with the PHAF, 
namely, 
the Heisenberg antiferromagnet (HAFM) 
on the simple-cubic lattice  
and 
on the kagome lattice.
The former one, that orders below the N\'{e}el temperature $T_N > 0$, 
can be considered as the unfrustrated counterpart of the PHAF, 
since the simple-cubic lattice has also six nearest neighbors. 
The latter one, that does not order in the ground state for low spin quantum number, 
can be considered as the two-dimensional analogue of the PHAF.

Most of the previous studies on the quantum PHAF were focused on the ground-state properties of the model.
Thus, a field-theory attempt to understand the nature of the ground state was reported in Ref.~\cite{Harris1991}.
The bond-operator-method calculations of Ref.~\cite{Isoda1998}
leads to a valence-bond-crystal state as the ground state of the model.
Perturbative expansions starting from noninteracting tetrahedral unit cells 
which yield the spin correlations for the model were performed in Refs.~\cite{Canals1998,Canals2000}. 
The conclusion of this study is that the ground state is a spin liquid with exponentially decaying correlations, 
where the correlation length does not exceed the nearest-neighbor distance.
Similar approaches starting from the limit of isolated tetrahedra and switching on the interactions between the tetrahedra as perturbation 
were later on presented in Refs.~\cite{Koga2001,Tsunetsugu2001,Tsunetsugu2001b,Tsunetsugu2017}.
The contractor renormalization method applied to the spin-1/2 PHAF
leads to the  conclusion that the ground state is a valence bond solid breaking lattice symmetry \cite{Berg2003}.
Other routes to the problem, 
which do not start from less symmetric Hamiltonians to be treated perturbatively,
were considered in Refs.~\cite{Moessner_2006,Tchernyshyov_2006}.
In these papers, the spin-1/2 problem on the pyrochlore lattice was studied 
after enlarging the symmetry of the spin space from ${\rm{SU}}(2)\sim {\rm{Sp}}(1)$ to ${\rm{Sp}}(N)$ \cite{Moessner_2006,Tchernyshyov_2006},
however, the large-$N$ physics cannot be uniquely transferred to ${\rm{SU}}(2)\sim {\rm{Sp}}(N=1)$ limit.
Fermionic mean-field theory followed by variational Monte Carlo \cite{Kim2008} 
as well as a large-$N$ SU($N$) fermionic mean-field theory \cite{Burnell2009}
suggested a chiral spin-liquid state as the ground state of the $S=1/2$ PHAF.
Large-$S$ approaches for the PHAF were discussed in Refs.~\cite{Henley_2006,Hizi_2007,Hizi_2009}.
They yield indications that via the {\it order by disorder mechanism} 
quantum fluctuations select collinear states among the huge degenerate manifold of classical ground states. 
We may mention here the difference to the kagome HAFM, 
where collinear states are not present in the classical ground-state manifold.

Among very recent papers on the quantum PHAF we may mention
an analytical study (a favor-wave theory combined with a mean-field approach) 
of a $S=1$  model with Dzyaloshinskii-Moriya interaction and single-ion spin anisotropy \cite{Pyro_S1_2017},
exact-diagonalization calculations for a $S=1/2$ system of up to 36 sites \cite{Chandra_ED_pyro_2018},
or
investigations of low-temperature phases of the quantum spin-$S$ PHAF including nearest-neighbor and next-nearest-neighbor interactions
using the pseudofermion functional renormalization group method (PFFRG) \cite{FPRG_Pyro_2018}.
The dynamical structure factor of the $S=1$ pyrochlore material NaCaNi$_2$F$_7$ 
has been studied with a combination of molecular dynamics simulations, stochastic dynamical theory and linear spin-wave theory \cite{Zhang2018}.

So far, less attention has been paid to finite-temperature properties.
We have to mention here the studies 
on the checkerboard lattice (planar pyrochlore) and pyrochlore-like models of the mineral clinoatacamite
using numerical linked-cluster expansions along with exact diagonalization of finite clusters \cite{Khatami2011,Khatami2012}.
Furthermore,
the diagrammatic Monte Carlo simulations for correlation functions down to the temperature $J/6$  were performed in Ref.~\cite{Huang2016}.
They reveal spin-ice states at $T=J/6$ although the lower temperatures remain inaccessible \cite{Huang2016}.
The above mentioned study \cite{FPRG_Pyro_2018} of the spin-$S$ $J_1-J_2$ Heisenberg model on the pyrochlore lattice employing the PFFRG 
includes both the ground-state and thermodynamic properties.
The theoretical study on the $S=1$ pyrochlore material NaCaNi$_2$F$_7$ \cite{Zhang2018} also refers to finite (although low) temperatures.
In what follows, we shall come back to some of these results. 

The main goal of the present study is to describe finite-temperature properties of the quantum PHAF.
In addition, 
we also present data for the ground-state energy, the uniform susceptibility,  
the excitation spectrum, the spin-spin correlation functions and structure factors at zero temperature.
The tool box to study finite-temperature  properties of the highly frustrated three-dimensional spin model is sparse.
Here we use two universal methods,
a second-order rotation-invariant Green's function method (RGM) \cite{Kondo1972}
and 
a high-temperature expansion (HTE) \cite{Oitmaa2006}.
While the HTE is restricted to temperatures above $\sim J$, the RGM is applicable for arbitrary temperatures.
  
The rest of the paper is organized as follows.
In Sec.~\ref{sec2} we briefly introduce the PHAF model and then in Sec.~\ref{sec3} we describe concisely the methods used for calculations.
We discuss our findings for the PHAF in Sec.~\ref{sec4} (zero-temperature results) and Sec.~\ref{sec5} (finite-temperature results).
Finally, in Sec.~\ref{sec6} we summarize our work.
The Appendix contains lengthy formulas for a few high-temperature-expansion terms
for the static structure factor of the PHAF with $S=1/2$, $1$, and $3/2$.

\section{Model}
\label{sec2}
\setcounter{equation}{0}

\begin{figure}
\begin{center}
\includegraphics[clip=on,width=80mm,angle=0]{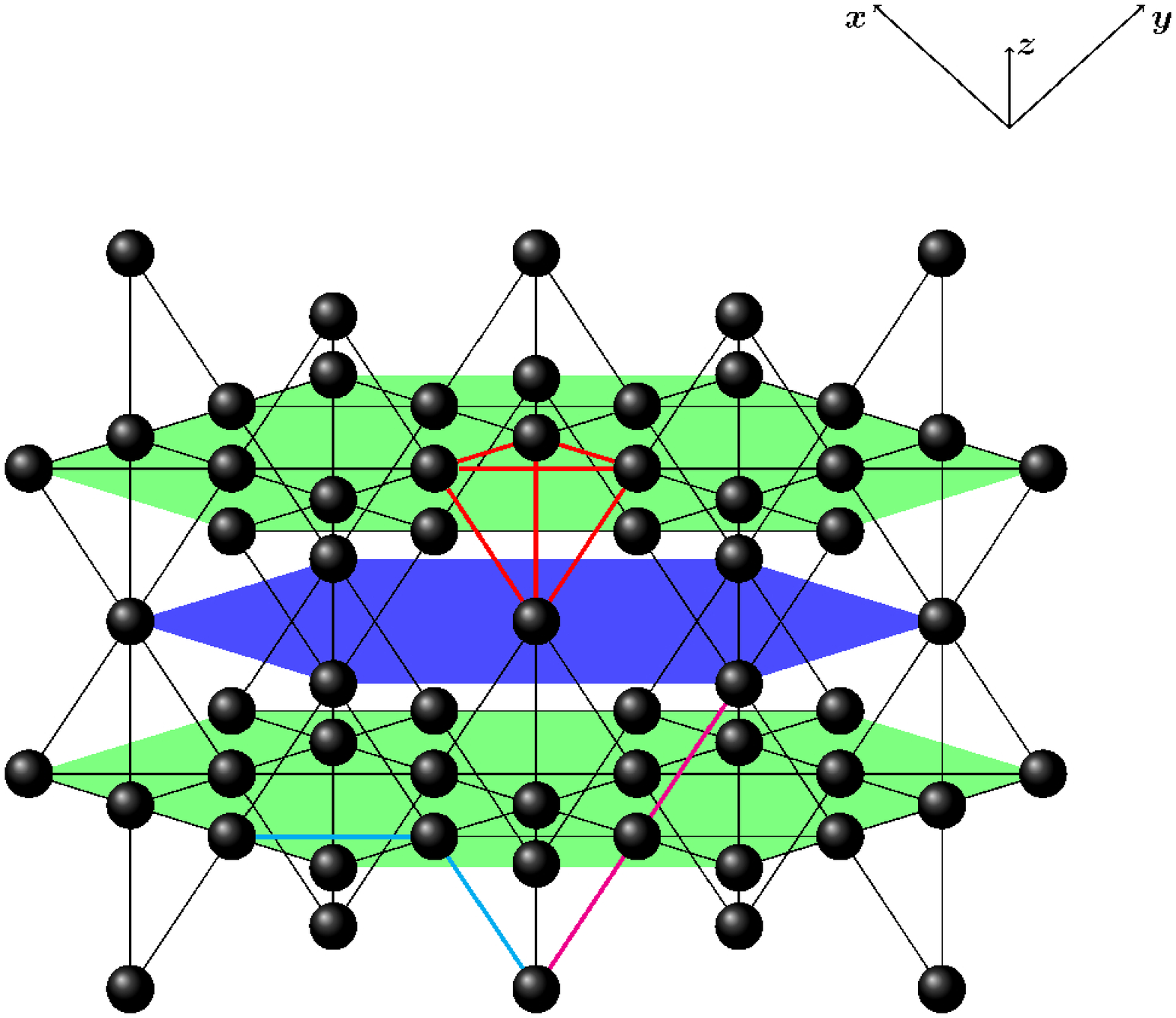}\\
\vspace{10mm}
\includegraphics[clip=on,width=65mm,angle=0]{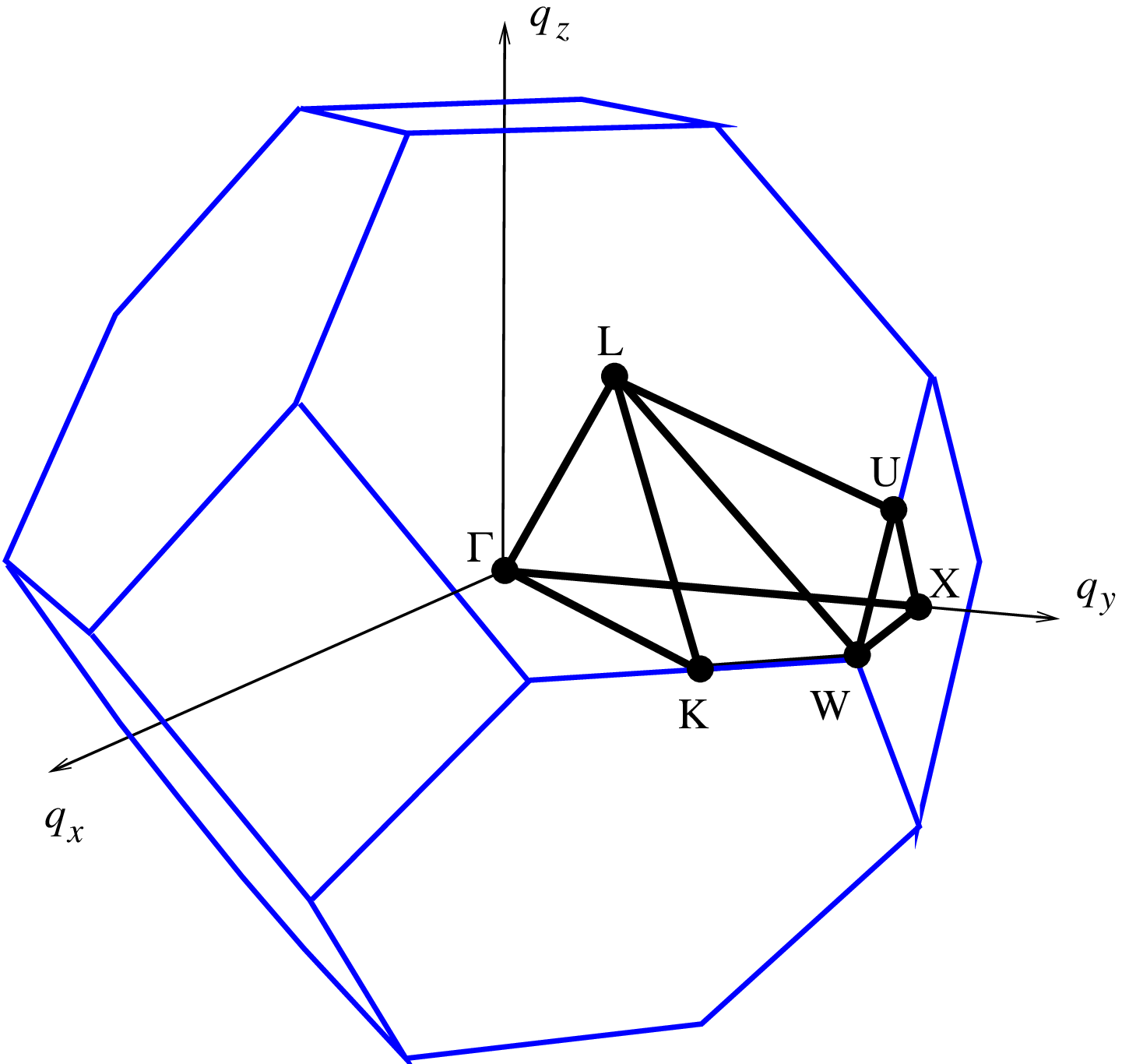}
\caption
{(Top)
The pyrochlore lattice 
visualized here as a three-dimensional structure which consists of alternating kagome (green) and triangular (blue) planar layers. 
The four-site unit cell is marked with the red bonds.
A red bond also indicates the nearest-neighbor correlation function $c_{100}$.
The path which connects the two sites entering the correlation function $c_{110}$ ($c_{200}$) is colored in cyan (violet), 
see the main text.
(Bottom)
The first Brillouin zone of a face-centered-cubic Bravais lattice.
The points $\Gamma$, X, W, K, U, and L in the ${\bf q}$-space are given by 
$\Gamma=(0,0,0)$, 
X$=(0,2\pi,0)$, 
W$=(\pi,2\pi,0)$, 
K$=(3\pi/2,3\pi/2,0)$,
U$=(\pi/2,2\pi,\pi/2)$,
and
L$=(\pi,\pi,\pi)$,
see, e.g., Refs.~\cite{symmetry_points,Burnell2009}.}
\label{fig01} 
\end{center}
\end{figure}

We consider the Heisenberg model on the pyrochlore lattice (see Fig.~\ref{fig01}, top) given by 
\begin{eqnarray}
\label{201}
\hat{H} = J\sum_{\langle m\alpha,n\beta\rangle}
\hat{{\bm{S}}}_{m\alpha}\cdot\hat{{\bm{S}}}_{n\beta}.
\end{eqnarray}
The sum in Eq.~(\ref{201}) runs over all nearest-neighbor bonds.
The antiferromagnetic nearest-neighbor coupling is set to unity, $J=1$, 
and $\hat{\boldsymbol{S}}_{m\alpha}^2=S(S+1)$, $S\ge 1/2$.
The pyrochlore lattice is described as four interpenetrating face-centered-cubic sublattices.
The origins of these sublattices are taken to be 
${\bf{r}}_1=(0,0,0)$,
${\bf{r}}_2=(0,1/4,1/4)$,
${\bf{r}}_3=(1/4,0,1/4)$,
and
${\bf{r}}_4=(1/4,1/4,0)$,
whereas the sites of the face-centered-cubic lattice are determined by 
${\bf{R}}_m=m_1{\bf{e}}_1+m_2{\bf{e}}_2+m_3{\bf{e}}_3$,
where $m_1$, $m_2$, $m_3$ are integers
and
${\bf{e}}_1=(0,1/2,1/2)$,
${\bf{e}}_2=(1/2,0,1/2)$,
${\bf{e}}_3=(1/2,1/2,0)$.
As a result,
the $N$ pyrochlore-lattice sites are labeled by $m\alpha$,
${\bf{R}}_{m\alpha}={\bf{R}}_m+{\bf{r}}_\alpha$,
where $m=1,\ldots, {\cal{N}}$, 
${\cal{N}}=N/4$ is the number of unit cells,
and $\alpha=1,2,3,4$ labels the sites in the unit cell.
The nearest-neighbor separation is $d=\sqrt{2}/4\approx 0.35$.
In Fig.~\ref{fig01} (bottom) we also show the first Brillouin zone of a face-centered-cubic Bravais lattice
along with some symmetric points in the ${\bf q}$-space to be used in what follows.

The pyrochlore-lattice Heisenberg Hamiltonian (\ref{201}) can be rewritten through a sum over $N/2$ tetrahedra \cite{Reimers1991,Moessner1998a,Moessner1998b}:
$2\hat{H}/J=\sum_{T=1}^{N/2}\hat{\boldsymbol{S}}_{T}^{2}-\sum_{T=1}^{N/2}\sum_{\alpha=1}^{4}\hat{\boldsymbol{S}}_{T\alpha}^{2}$,
where 
$\hat{\boldsymbol{S}}_{T}=\hat{\boldsymbol{S}}_{T1}+\hat{\boldsymbol{S}}_{T2}+\hat{\boldsymbol{S}}_{T3}+\hat{\boldsymbol{S}}_{T4}$
is the total spin of the tetrahedron $T$ and $\hat{\boldsymbol{S}}_{T\alpha}^{2}= S(S+1)$. 
In the classical limit $S\to\infty$, 
when all $\hat{\boldsymbol{S}}^2_{T}$ commute,
the ground-state configurations are given by the constraint $\hat{\boldsymbol{S}}^2_{T}=0$ on each tetrahedron separately. 
This results in a massive ground-state degeneracy, 
although the ground-state energy per site is quite simple and is given by $E_0/N = -S^{2}J$.

From the experimental side, there are only a few compounds which can be described by the model (\ref{201}). 
In addition to the already mentioned fluoride NaCaNi$_2$F$_7$  
which provides a good realization of the $S = 1$ PHAF, 
there are compounds which at least in their high-temperature phases are candidates for the PHAF (\ref{201}). 
For example, the molybdate Y$_2$Mo$_2$O$_7$
(which, however, shows spin-glass behavior at low temperature and spin-orbit coupling is relevant)
\cite{Greedan1986,Silverstein2014,Thygesen2017},
the chromites $A$Cr$_2$O$_4$ ($A$=Mg,Zn,Cd) 
(which, however, show a magneto-structural transition at low temperatures) 
\cite{Gao2018,Ji2009,Matsuda2007}, 
or
FeF$_3$ 
(for which besides the nearest-neighbor antiferromagnetic Heisenberg interaction, 
also biquadratic and Dzyaloshinskii-Moriya interactions are present) 
\cite{Sadeghi2015}.

\section{Methods}
\label{sec3}
\setcounter{equation}{0}

\subsection{Rotation-invariant Green's function method (RGM)}
\label{sec3A}

Our first method used in the present study of the PHAF is a double-time temperature-dependent Green's function technique
which is widely employed in quantum many-body theory \cite{Tyablikov1967,Gasser2001,Froebrich2006}.
An important development of this approach was achieved by Kondo and Yamaji in 1972 \cite{Kondo1972}
by introducing a rotation-invariant formalism to describe short-range order of the one-dimensional $S=1/2$ Heisenberg ferromagnet at $T>0$.
Going one step beyond the usual random-phase approximation (Tyablikov approximation) \cite{Tyablikov1967,Gasser2001,Froebrich2006,Nolting2009,Hutak2018}
the rotational invariance is introduced by setting $\langle \hat{S}^z_{i}\rangle=0$ in the equations of motion. 
Within this scheme possible magnetic long-range is described by the long-range term (condensation part) in the spin-spin correlation function, 
see, e.g., Refs.~\cite{Shimahara1991,Ihle1997,Siurakshina2000,Siurakshina2001}.
Moreover, the decoupling approximation of higher-order correlators is improved   
by introducing so-called vertex parameters, see below.
We mention here that the first-order random-phase approximation fails for the PHAF, 
since it is not appropriate to describe magnetic phases with short-range order \cite{Kondo1972,Ihle1997,Haertel2011a,Mueller2017a,Mueller2017b,Hutak2018}.

Since 1972 the rotation-invariant Green's function method (RGM) was continuously further developed 
and nowadays it is a well-established technique that has been used in numerous recent studies on quantum spin systems
(including arbitrary quantum spin number $S$, any lattice dimension, lattices with non-primitive unit cell, geometrically frustrated lattices)
\cite{Yu2000,Siurakshina2000,Siurakshina2001,Bernhard2002a,junger2004green,Junger2005,Junger2009,Schmalfus2004a,Schmalfus2005a,Schmalfus2006,
Haertel2010,Haertel2011a,Haertel2011b,Haertel2013,Antsygina2008,Antsygina2009,Mikheyenkov2013,Mikheyenkov2016,Vladimirov2014,Vladimirov2017,
Mueller2015,Mueller2017a,Mueller2017b,Mueller2018}.

To be more specific, 
in the present study of the PHAF we deal with a set of Green's functions 
$\langle \langle \hat{S}^\mu_{{\bf{q}}\alpha};\hat{S}^\nu_{{\bf{q}}\beta}\rangle\rangle_\omega
=-\chi_{{\bf{q}}\alpha\beta}^{\mu\nu}(\omega)$,
where 
$\langle\langle \hat{A};\hat{B}\rangle\rangle_t
=-{\rm{i}}\Theta(t)\langle [\hat{A}(t), \hat{B}]_-\rangle$,
the subscript $\omega$ means the Fourier transform with respect to the time $t$,
$\mu\nu$ denotes $+-$ or $zz$,
and
$\hat{S}^+_{{\bf{q}}\alpha}=\sum_m\exp(-{\rm{i}}{\bf{q}}\cdot{\bf{R}}_m) \hat{S}^+_{m\alpha}/\sqrt{{\cal{N}}}$
(the sum runs over all unit cells, i.e., $m=1,\ldots,{\cal{N}}$).
Moreover, the dynamical susceptibilities $\chi_{{\bf{q}}\alpha\beta}^{\mu\nu}(\omega)$ are immediately known once the Green's functions are determined.

In Ref.~\cite{Mueller2017b} it was shown that within the framework of the RGM the equations of motion can be compactly written in the following matrix form:
\begin{eqnarray}
\label{301}
(\omega^2 I - F_{\mathbf{q}})\chi^{{+-}}_{\mathbf{q}}(\omega) = -M_{\mathbf{q}}.
\end{eqnarray}
Since the unit cells contains four sites, the matrices in Eq.~\eqref{301} are $4\times 4$ Hermitian matrices, 
namely, 
the unit matrix $I$,
the frequency matrix $F_\mathbf{q}$, 
the susceptibility matrix $\chi^{{+-}}_\mathbf{q}(\omega)$,
and
the moment matrix $M_\mathbf{q}$.
Although the study of Ref.~\cite{Mueller2017b} concerns the spin-$S$ ferromagnetic case,
the RGM equations derived there hold for the antiferromagnetic coupling $J=1$, too,
because they do not depend on the sign of the exchange interaction.
(For explicit expressions for the moment matrix and the frequency matrix see Eqs.~(5) and (6) in Ref.~\cite{Mueller2017b}.)
Importantly,
these matrix elements contain spin correlation functions 
$c_{ijk} \equiv \langle \hat{S}^+_{\mathbf{0}} \hat{S}^-_{\mathbf{R}}\rangle$,
${\mathbf{R}}=i{\bf{r}}_2+j{\bf{r}}_3+k{\bf{r}}_4$. 
Due to lattice symmetry, 
only the non-equivalent correlators $c_{100}$, $c_{110}$, and $c_{200}$ enter the matrix elements,
where $c_{100}$ is related to the sites connected by the edge of unit-cell tetrahedron
(nearest-neighbor correlator),
$c_{110}$ is related to the sites of two adjacent unit cell tetrahedra connected by two noncollinear edges with a common site 
(next-nearest-neighbor correlator),
and 
$c_{200}$ is related to the sites of two adjacent unit cell tetrahedra connected by two collinear edges with a common site
(one of the two kinds of third-neighbor correlators),
see Fig.~\ref{fig01}, top.
These correlators appear in the matrix elements through $\tilde{\alpha}_{ijk}=\alpha_{ijk}c_{ijk}$ and $\tilde{\lambda}_{ijk}=\lambda_{ijk}c_{ijk}$.
Here $\alpha_{ijk}$ and $\lambda_{ijk}$ are the vertex parameters 
which are introduced to improve the approximation made by the decoupling in second order,
e.g.,
$\hat{S}^+_{A}\hat{S}^-_{B}\hat{S}^z_{C}\to \alpha_{AB}c_{AB}^{+-} \hat{S}^z_{C}$
or
$\hat{S}^+_{A}\hat{S}^-_{B}\hat{S}^z_{B}\to \lambda_{AB}c_{AB}^{+-} \hat{S}^z_{B}$.
Moreover, we have $\lambda_{ijk}=0$ for $S=1/2$.

Going back to Eq.~(\ref{301}),
it is important to note that the moment matrix $M_\mathbf{q}$ and the frequency matrix $F_\mathbf{q}$ commute,
i.e., $[M_\mathbf{q},F_\mathbf{q}]_- = 0$.
Let us denote the common eigenvectors of the matrices $M_\mathbf{q}$ and $F_\mathbf{q}$ 
by $\vert{\gamma\mathbf{q}}\rangle$, $\gamma=1,2,3,4$.
Moreover, let us introduce their eigenvalues,
i.e.,
$M_{\mathbf{q}}|{\gamma\mathbf{q}}\rangle=m_{\gamma\mathbf{q}}|{\gamma\mathbf{q}}\rangle$
and
$F_{\mathbf{q}}|{\gamma\mathbf{q}}\rangle=\omega^2_{\gamma\mathbf{q}}|{\gamma\mathbf{q}}\rangle$.
In Ref.~\cite{Mueller2017b} it has been found that
\begin{eqnarray}
\label{302} 
\frac{m_{1\mathbf{q}}}{J} = \frac{m_{2\mathbf{q}}}{J} = \frac{m_{3\mathbf{q}}}{J}+\frac{m_{4\mathbf{q}}}{J} & = & -16 c_{100},
\nonumber\\
\frac{m_{3\mathbf{q}}}{J}-\frac{m_{4\mathbf{q}}}{J} & = & -8 c_{100}D_{\mathbf{q}}
\end{eqnarray}
with
$D_{\mathbf{q}}^2 = 1 +\cos(q_x/2)\cos(q_y/2) +\cos(q_x/2)\cos(q_z/2) +\cos(q_y/2)\cos(q_z/2)$
and
\begin{eqnarray}
\label{303} 
\frac{\omega^2_{1\mathbf{q}}}{J^2}\! =\!\frac{\omega^2_{2\mathbf{q}}}{J^2}
&=&\frac{8}{3}(2S(S+1)+3\tilde{\lambda}_{100}
\\
&+& 9\tilde{\alpha}_{100} + 6 \tilde{\alpha}_{110}+3 \tilde{\alpha}_{200}),
\nonumber\\
\frac{\omega^2_{3\mathbf{q}}}{J^2}\!+\!\frac{\omega^2_{4\mathbf{q}}}{J^2} 
&=& \frac{8}{3}(2S(S+1)+3\tilde{\lambda}_{100}
\nonumber\\
&+& 3(D_{\mathbf{q}}^2-1)\tilde{\alpha}_{100}+6\tilde{\alpha}_{110}+3\tilde{\alpha}_{200}),
\nonumber\\
\frac{\omega^2_{3\mathbf{q}}}{J^2}\!-\!\frac{\omega^2_{4\mathbf{q}}}{J^2} 
&=& \frac{8}{3}D_{\mathbf{q}}S(S+1)
\nonumber\\
&+& 4 D_{\mathbf{q}} (\tilde{\lambda}_{100}+3\tilde{\alpha}_{100}+2\tilde{\alpha}_{110}+\tilde{\alpha}_{200}). 
\nonumber
\end{eqnarray}
The common eigenvectors $\vert \gamma{\bf{q}}\rangle$ of the moment matrix $M_\mathbf{q}$ and the frequency matrix $F_\mathbf{q}$ are rather lengthy; 
they are presented in Appendix~B in Ref.~\cite{Mueller2017b}.

Now we can resolve Eq.~(\ref{301}) to find the set of  dynamical susceptibilities (Green's functions).
They are given by
\begin{eqnarray}
\label{304}
\chi^{+-}_\mathbf{q\alpha\beta}(\omega) 
&=& -\sum_{\gamma} \frac{m_{\gamma\mathbf{q}}}{\omega^2-\omega^2_{\gamma\mathbf{q}}}\langle\alpha|{\gamma\mathbf{q}}\rangle\langle{\gamma\mathbf{q}}|\beta\rangle,
\end{eqnarray}
where $\langle\alpha|{\gamma\mathbf{q}}\rangle$ is the $\alpha$th component of the eigenvector $|{\gamma\mathbf{q}}\rangle$.
The correlation functions are obtained by applying the spectral theorem
\begin{eqnarray}
\label{305}
c_{m\alpha,n\beta} 
&=& \frac{1}{\mathcal{N}}\sum_{\mathbf{q}\ne\mathbf{Q}}c_{\mathbf{q}\alpha\beta}\cos(\mathbf{q}\cdot\mathbf{r}_{m\alpha,n\beta}) 
\nonumber \\
&+&\sum_{\mathbf{Q}}C_{\mathbf{Q}\alpha\beta}\cos(\mathbf{Q}\cdot\mathbf{r}_{m\alpha,n\beta})
\end{eqnarray}
with
\begin{eqnarray}
\label{306}
c_{\mathbf{q}\alpha\beta}
&=&\sum_{\gamma}\frac{m_{{\gamma}{\bf{q}}}}{2\omega_{\gamma\mathbf{q}}}(1+2n(\omega_{\gamma\mathbf{q}}))\langle\alpha|{\gamma\mathbf{q}}\rangle\langle{\gamma\mathbf{q}}|\beta\rangle,
\end{eqnarray}
where 
$n(\omega)=1/(\exp(\omega/T)-1)$ is the Bose-Einstein distribution function 
and 
$C_{\mathbf{Q}\alpha\beta}$ is the so-called condensation term which is related to magnetic long-range order, 
see, e.g., Refs.~\cite{Shimahara1991,Ihle1997,Siurakshina2000,Siurakshina2001}.
For example, 
for the ferromagnet \cite{Mueller2017b},
only one condensation term at $\mathbf{Q}=\mathbf{0}$ is relevant, 
i.e., $C_{\mathbf{0}\alpha\beta}=C_{\mathbf{0}}$, 
and the total magnetization is given by the expression $M=\sqrt{3C_{\mathbf{0}}/2}$. 
The susceptibility $\chi_{\mathbf{Q}}$ is given by the expression
\begin{eqnarray}
\label{307}
\chi_{\mathbf{Q}} &\equiv& \chi^{zz}_{\mathbf{Q}}=\chi^{+-}_{\mathbf{Q}}/2
=
\underset{(\mathbf{q},\omega)\rightarrow(\mathbf{Q},0)}{\textrm{lim}} 
\frac{1}{4}\sum_{\alpha}\sum_{\beta}\frac{\chi_{\mathbf{q}\alpha\beta}^{+-}(\omega)}{2}
\nonumber\\
&=&
\underset{\mathbf{q}\rightarrow\mathbf{Q}}{\textrm{lim}}\sum_{\alpha,\beta,\gamma}\frac{m_{\gamma\mathbf{q}}}{8\omega^2_{\gamma\mathbf{q}}}
\langle\alpha\vert\gamma{\bf{q}}\rangle\langle \gamma{\bf{q}}\vert\beta\rangle.
\end{eqnarray}
In case of magnetic long-range order,  
$\chi_{\mathbf{Q}}$ diverges at a critical temperature $T_c$, where ${\bf{Q}}$ is the magnetic wave vector.
According to Eq.~\eqref{307},
this would be related to divergence of $m_{\gamma\mathbf{q}}/\omega^2_{\gamma\mathbf{q}}$ as $\mathbf{q}\rightarrow \mathbf{Q}$.
For the further discussion of the relevant RGM equations it is important to state here, 
that within the RGM for the PHAF we do not find such a divergence for all $\mathbf{Q}$ and all temperatures $T\ge 0$.
This means that for the PHAF there is no condensation term or, in other words, no magnetic long-range order for all temperatures $T\ge 0$.

Knowing the dynamical susceptibilities or the Green's functions (\ref{304}) and the correlation functions (\ref{305}), (\ref{306}),
we can easily obtain the (zero-frequency) susceptibility $\chi_{\mathbf{Q}}$ (\ref{307}) and the specific heat $C_V$.
Furthermore, 
using Eq.~(\ref{306}) we can also obtain the static structure factor 
$S_{\mathbf{q}}=3S_{\mathbf{q}}^{+-}/2$,
$S_{\bf{q}}^{+-}=\sum_{\alpha,\beta}c_{\mathbf{q}\alpha\beta}/4$.
Last but not least,
the dynamical structure factor 
$S^{zz}_{{\bf{q}}}(\omega)=S^{+-}_{{\bf{q}}}(\omega)/2$ 
follows from the fluctuation-dissipation theorem,
i.e.,
$S^{+-}_{{\bf{q}}}(\omega)=(2/(1-e^{-\omega/T}))\Im\chi^{+-}_{{\bf{q}}}(\omega)$,
$\chi^{+-}_{{\bf{q}}}(\omega)=\sum_{\alpha,\beta}\chi^{+-}_{{\bf{q}}\alpha\beta}(\omega)/4$.
Thus, Eq.~(\ref{304}) leads straightforwardly to $S^{+-}_{{\bf{q}}}(\omega)$.
After some standard manipulations we arrive at
\begin{eqnarray}
\label{308}
S_{{\bf{q}}}^{zz}(\omega)
=
\frac{\pi}{1-e^{-\frac{\omega}{T}}}\sum_{\alpha,\beta}\sum_{\gamma}\frac{m_{\gamma{\bf{q}}}}{8\omega_{\gamma{\bf{q}}}}
\nonumber\\
\times
\left(
\delta(\omega-\omega_{\gamma{\bf{q}}})
-
\delta(\omega+\omega_{\gamma{\bf{q}}})
\right)
\langle \alpha\vert\gamma{\bf{q}}\rangle \langle \gamma{\bf{q}}\vert\beta\rangle.
\end{eqnarray}
This quantity is related to neutron inelastic scattering data accessible in experiments.
We also note that integrating $S_{{\bf{q}}}^{zz}(\omega)$ (\ref{308}) over all $\omega$ we get the static structure factor:
\begin{eqnarray}
\label{309}
\int_{-\infty}^{\infty}{\rm{d}}\omega S_{{\bf{q}}}^{zz}(\omega)
=
2\pi S_{{\bf{q}}}^{zz}
=
2\pi\frac{1}{3}S_{{\bf{q}}}.
\end{eqnarray}
Note that there is no intrinsic damping within the RGM approach.
Therefore, we replace the $\delta$-functions in Eq.~(\ref{308}) by the Lorentzian function, 
i.e., $\delta(x)\to (1/\pi)(\epsilon/(x^2+\epsilon^2))$, 
where the ``damping'' parameter $\epsilon$ is chosen as $\epsilon = 0.1$.

In summary,
for the considered antiferromagnetic case, i.e., $J=1$,
we have to solve self-consistently the equations for the correlation functions $c_{100}$, $c_{110}$, $c_{200}$, and the vertex parameters. 
Taking into account all possible vertex parameters $\alpha_{ijk}(T)$ and $\lambda_{ijk}(T)$ would therefore exceed the number of available equations.
In the simplest version of the RGM, often called the minimal version,
one considers only one vertex parameter in each class,
i.e., $\alpha_{ijk}(T)=\alpha(T)$ and $\lambda_{ijk}(T)=\lambda(T)$. 
We mention, 
that this simple version with only one $\alpha$ parameter  
was used in the early RGM kagome papers for the $S=1/2$ case (where $\lambda(T)\equiv0$) \cite{Yu2000,Bernhard2002a,Schmalfus2004a}.
An improvement of the minimal version can be achieved by taking into account more vertex parameters, 
however, that requires additional input to get more equations for the additional vertex parameters. 
For example, in Ref.~\cite{Mueller2018}, 
for the kagome-lattice spin-$S$ HAFM,
two $\alpha$ parameters 
($\alpha_1$ for nearest-neighbor sites and $\alpha_2$ for not-nearest-neighbor sites) 
are introduced 
and the value of the ground-state energy obtained by the coupled cluster method (CCM) \cite{Goetze2011,Goetze2015} is used as an additional input.
In the case of the quantum PHAF we do not have such data,
and, therefore, we have to restrict ourselves to the minimal version of the RGM.
In Ref.~\cite{Mueller2018}, by comparison of the minimal and the extended version using the CCM input, 
it has been found that for the kagome HAFM the minimal version works reasonably well for small spin quantum numbers $S$, 
but may fail for large $S$.
	
Within the minimal version the set of equations is found as follows.
For every unknown correlation function the spectral theorem yields one equation.
One more equation is given by the sum rule $\bm{\hat{S}}_{m\alpha}^2=S(S+1)$, which determines, e.g., one vertex parameter.
Thus, for $S=1/2$, where $\lambda=0$, these equations determine all unknown quantities.
For $S>1/2$, additionally the unknown parameter $\lambda$ has to be determined. 
For that we follow Refs.~\cite{Junger2005,Junger2009,Haertel2011a,Vladimirov2014,Mueller2015,Mueller2017b,Vladimirov2017,Mueller2018}.
At zero temperature we use the well-tested ansatz $\lambda(0)=2-1/S$.
At infinite temperature $\lambda(\infty)=1-3/(4S(S+1))$ is valid,
as it has been verified by comparison with the high-temperature expansion, see, e.g., \cite{Junger2005}.
For intermediate temperatures we set the ratio
\begin{eqnarray}
\label{310}
r(T)\equiv\frac{\lambda(T)-\lambda(\infty)}{\alpha(T)-\alpha(\infty)}
=\frac{\lambda(0)-\lambda(\infty)}{\alpha(0)-\alpha(\infty)}
\end{eqnarray}
as temperature independent.

\subsection{High-temperature expansion (HTE)}
\label{sec3B}

Our second method used in the present study of the PHAF is the high-temperature expansion (HTE) 
which is a universal and straightforward approach in the theory of spin systems \cite{Oitmaa2006}.
To be more specific,
in the present study we use the HTE program of Ref.~\cite{Lohmann2014}, 
which is freely available at \verb"http://www.uni-magdeburg.de/jschulen/HTE/",
in an extended version up to 13th (11th) order for $S=1/2$ ($S>1/2$).
With this tool,
we compute the series of the static uniform susceptibility $\chi_{\mathbf{0}}=\sum_nc_n\beta^n$ and the specific heat $C_V=\sum_n d_n\beta^n$ 
with respect to the inverse temperature $\beta=1/T$. 
To extend the region of validity of the power series we use Pad\'{e} approximants denoted by $[m,n]=P_m(\beta)/Q_n(\beta)$,
where $P_m(\beta)$ and $Q_n(\beta)$ are polynomials in $\beta$ of order $m$ and $n$, respectively.
The coefficients of $P_m(\beta)$ and $Q_n(\beta)$ are determined by the condition 
that the expansion of $[m,n]$ has to agree with the initial power series up to order ${\cal{O}}(\beta^{m+n})$.

In addition, 
the high-temperature series of the static spin pair correlation function $\langle \hat{{\bf{S}}}_i\cdot\hat{{\bf{S}}}_j\rangle$ 
are calculated up to 12th order of $\beta$ (for $S=1/2$) or 10th order of $\beta$ (for $S>1/2$),   
following the strategy of Refs.~\cite{Lohmann2011,Lohmann2014,Richter2015}. 
Having the series of the correlation functions we evaluate the magnetic static structure factor 
\begin{eqnarray}
\label{311}
S_{\mathbf{q}}
=
\frac{1}{N}\sum_{i,j}\langle \hat{{\bf{S}}}_i\cdot\hat{{\bf{S}}}_j\rangle\cos(\mathbf{q}\cdot(\mathbf{R}_{i}-\mathbf{R}_{j})),  
\end{eqnarray}
see, e.g., Refs.~\cite{Richter2015,Mueller2018}.
Here $i$ and $j$ are the sites of the pyrochlore lattice labeled in Sec.~\ref{sec2} by $m\alpha$.
Evidently, $S_{\mathbf{q}}=3S_{\mathbf{q}}^{+-}/2$.

\section{Zero-temperature properties}
\label{sec4}
\setcounter{equation}{0}

\begin{figure}
\centering 
\includegraphics[clip=on,width=80mm,angle=0]{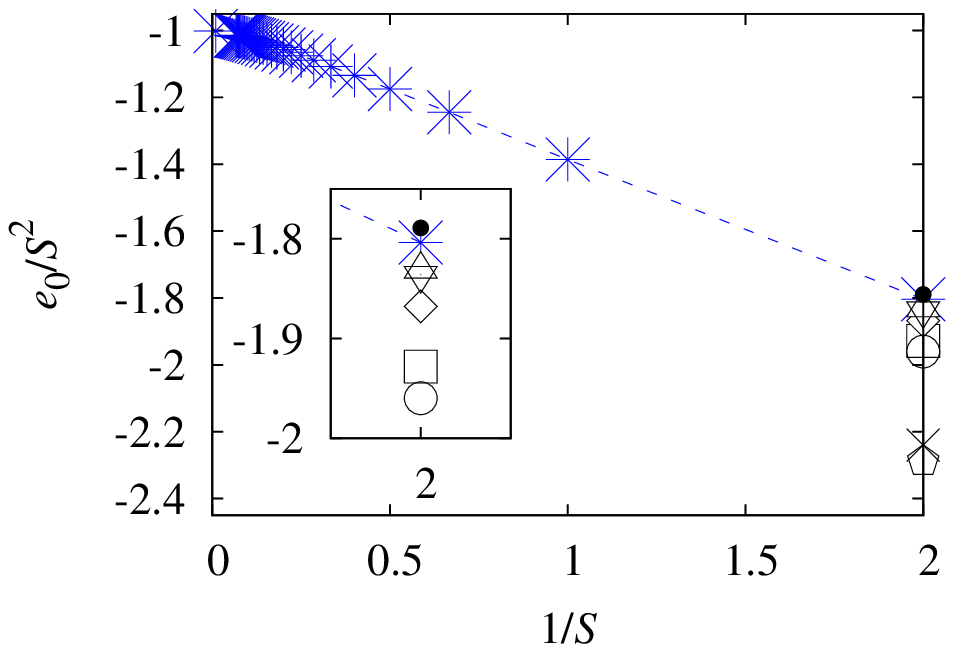}\\
\vspace{3mm}
\includegraphics[clip=on,width=80mm,angle=0]{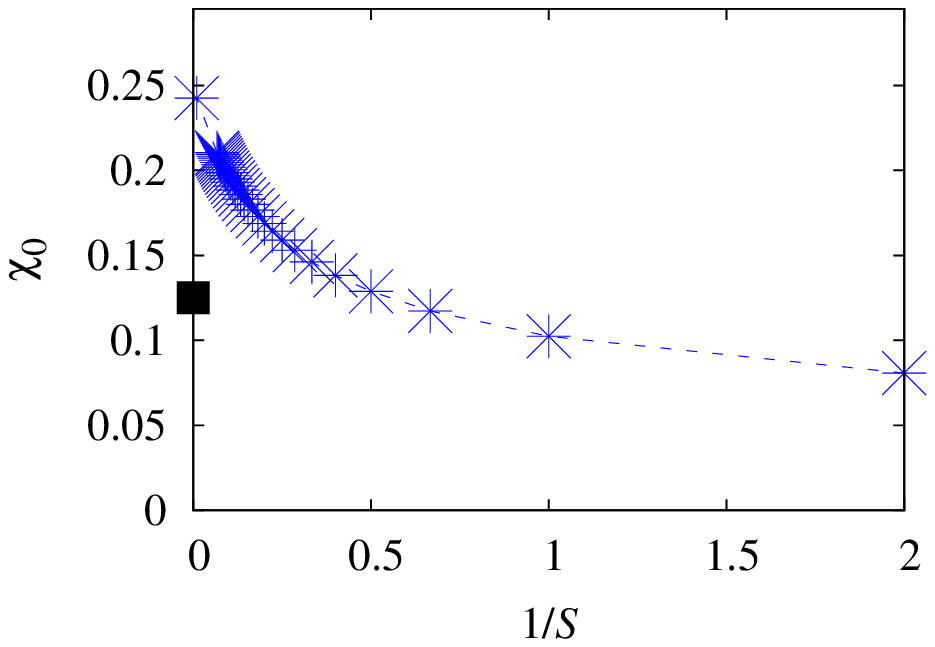}
\protect
\caption
{RGM results (blue symbols) 
for the ground-state energy $E_{0}/(NS^{2})$ (top) and the ground-state uniform susceptibility $\chi_{\bf{0}}$ (bottom)
of the PHAF ($J=1$)
as a function of the inverse spin-quantum number $1/S$.
The black symbols in the upper panel correspond to the results of 
Ref.~\cite{Burnell2009} (filled circles),
Refs.~\cite{Harris1991,Koga2001} (open circles),
Ref.~\cite{Sobral1997} (pentagons),
Ref.~\cite{Isoda1998} (up-triangles), 
Ref.~\cite{Canals2000} (crosses),
Ref.~\cite{Kim2008} (down-triangles);
squares and diamonds correspond to exact-diagonalization data for $N=28$ and $N=36$, respectively \cite{Chandra_ED_pyro_2018}.
The black square in the lower panel corresponds to the result of classical Monte Carlo simulations \cite{Moessner1999}.}
\label{fig02} 
\end{figure}

We begin this section with a discussion of the quality of the minimal-version RGM for the PHAF. 
As briefly explained in Sec.~\ref{sec3A}, 
the minimal version neglects the real-space dependence of the $\alpha$ parameter and is believed to be justified preferably for ferromagnets.
To estimate the accuracy of the adopted scheme for the PHAF we follow Ref.~\cite{Mueller2018}
and consider the RGM ground-state energy as well as the ground-state uniform susceptibility $\chi_{\bf{0}}$ as a function of $1/S$,  
see Fig.~\ref{fig02}.
It is obvious that in the classical limit $S\to\infty$ we obtain the correct result for the ground-state energy $E_0=-NS^2J$ \cite{Reimers1991}
(Fig.~\ref{fig02}, top).
Note that this is contrary to the case of the kagome-lattice HAFM, 
where the minimal version in the classical limit $S\to\infty$ gives a higher energy value than the exact one
and the discrepancy was removed after adopting the extended version \cite{Mueller2018}.
The ground-state energies per site for the pure quantum case of $S=1/2$ obtained by other approaches 
exhibit a pretty wide distribution (see the black symbols in Fig.~\ref{fig02}, top) 
ranging from $e_0\approx -0.572 J$ \cite{Sobral1997} to $e_0\approx -0.447 J$ \cite{Burnell2009},
thus providing evidence that reliable values in this limit are still lacking.
The ground-state uniform susceptibility $\chi_{\bf{0}}$ is shown in the lower panel of Fig.~\ref{fig02}.
As a function of the inverse spin quantum number 
$\chi_{\bf{0}}$ exhibits a noticeable upturn for $S \gtrsim 2$ 
leading finally to a significant deviation from classical Monte-Carlo result $\chi_{\bf{0}}\approx 0.125$ \cite{Moessner1999,PhysRevB.63.140404,PhysRevB.65.184418}. 
Note here that the kagome HAFM exhibits an unphysical divergence of the ground-state value of $\chi_{\bf{0}}$ as $S\to \infty$ 
when using the minimal version of the RGM \cite{Mueller2018}.
Thus, we may conclude, that the minimal version of the RGM likely works reasonably well for the ground state of the PHAF, 
however, for increasing $S$ the RGM data become less reliable.

\begin{figure}
\centering 
\includegraphics[clip=on,width=80mm,angle=0]{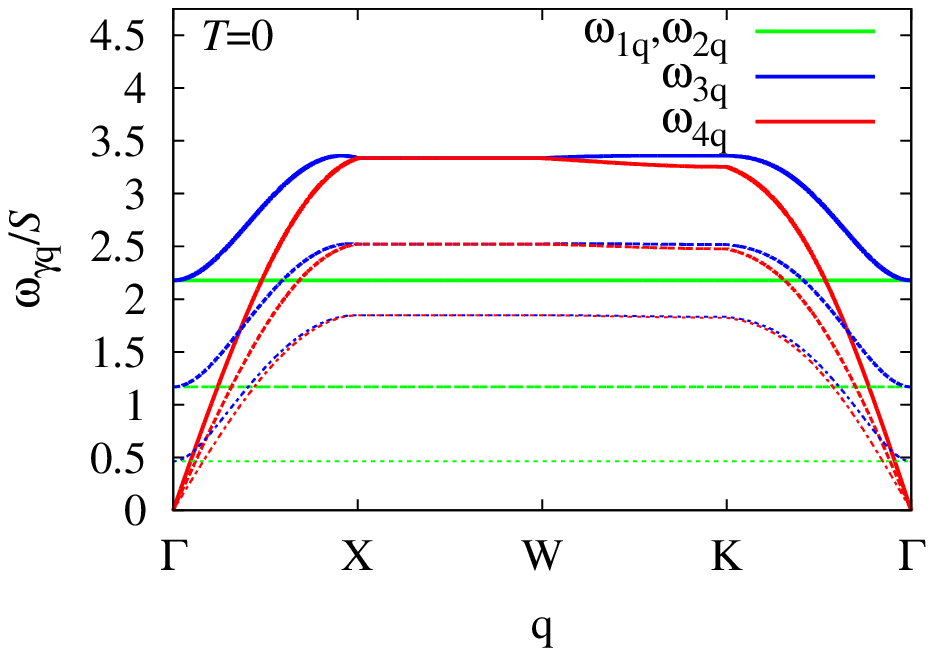}\\
\vspace{3mm}
\includegraphics[clip=on,width=80mm,angle=0]{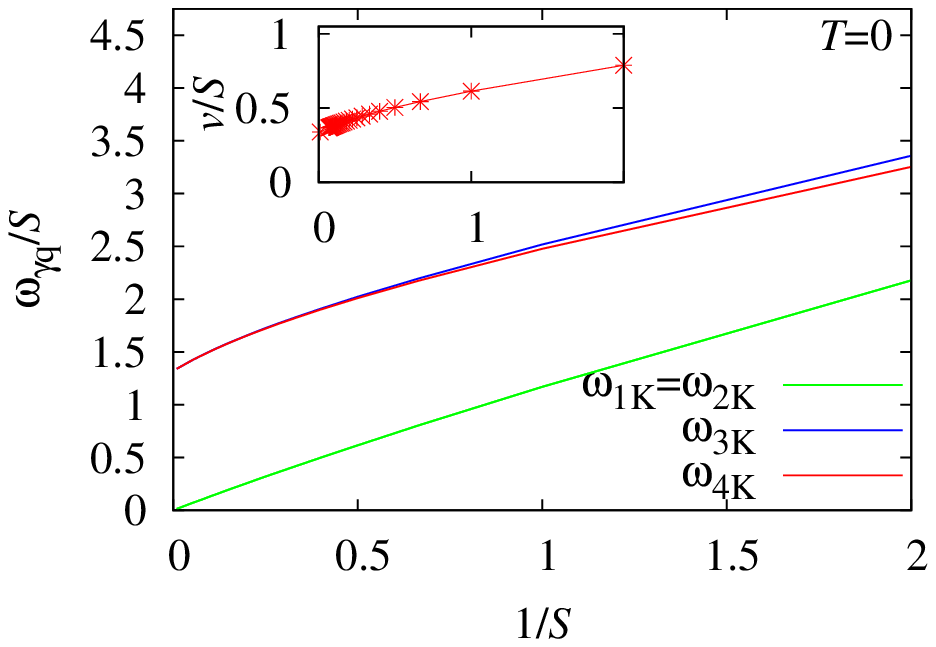}
\protect
\caption
{(Top)
Dispersion of the excitation energies $\omega_{\gamma\mathbf{q}}/S$
(Eq.~(\ref{303}), $J=1$)
at zero temperature $T=0$ for $S=1/2$ (thick), $S=1$ (thin), and $S=3$ (very thin).
The points $\Gamma$, X, W, and K in the first Brillouin zone of a face-centered-cubic Bravais lattice 
are given by $\Gamma=(0,0,0)$, X$=(0,2\pi,0)$, W$=(\pi,2\pi,0)$, and K$=(3\pi/2,3\pi/2,0)$, 
see Fig.~\ref{fig01}, bottom.
(Bottom, main panel) 
The ground-state excitation energies $\omega_{\gamma\mathbf{q}}/S$ in dependence on the inverse spin quantum number $1/S$ 
at ${\bf{q}}=(3\pi/2,3\pi/2,0)$ (K point). 
(Bottom, inset)
Normalized RGM ground-state excitation velocity $v/S$ in dependence on $1/S$.}
\label{fig03} 
\end{figure}

We turn to the discussion of the ground-state excitation spectrum for the PHAF.
We start with a brief discussion of the linear-spin-wave spectrum \cite{Sobral1997}.
The starting point of the linear spin-wave theory is a classical ground state. 
In the case of the PHAF the classical ground state has a huge degeneracy.
In Ref.~\cite{Sobral1997}, 
several classical ordered ground states with identical magnetic and crystallographic unit cells were considered (so-called ${\bf q}={\bf 0}$ states).
In all considered cases the linear-spin-wave spectrum contains flat zero-energy as well as dispersive modes.
In particular,
for the collinear classical state there are two degenerate flat zero-energy modes and two degenerate dispersive modes;
for the noncollinear ground state, 
where the spins point along the diagonals of the tetrahedron, 
all four modes are different and the lowest one is the flat zero-energy mode.

The RGM data for the excitation spectrum  $S=1/2$ (thick), $S=1$ (thin), and  $S=3$ (very thin) are shown in the upper panel of Fig.~\ref{fig03}.
Within the RGM we do not start from a peculiar classical ground state.
Moreover, the numerical computation of the spectrum has to be performed for each $S$ value separately.
As a result, we get $S$-dependent excitations $\omega_{\gamma\mathbf{q}}/S$, as we should expect using a more sophisticated approach. 
(Note that for the pyrochlore-lattice quantum Heisenberg ferromagnet 
the ground-state excitations energies $\omega_{\gamma\mathbf{q}}/S$ do not depend on $S$, 
since the ground state is classical \cite{Mueller2017b}.)
For finite $S$ the differences to the linear-spin-wave spectrum of \cite{Sobral1997} are obvious:
The flat (dispersionless) branch (green) is not the lowest one.
It is two-fold degenerate (as that of linear spin-wave theory for the collinear state) 
and its energy tends to zero as $S$ increases (Fig.~\ref{fig03}, lower panel) thus approaching the linear-spin-wave result.
There are also two dispersive branches, one is gapless (red) and one is gapped  (blue), which approach each other as $S$ increases, 
i.e., again linear-spin-wave result is obtained for $S\to\infty$ (Fig.~\ref{fig03}, lower panel).
Apparently, the RGM decoupling procedure (that is not biased in favor of a classical ground state) 
is in favor of linear spin-wave theory starting from the collinear classical state \cite{Sobral1997}, 
but not necessarily a ${\bf q}={\bf 0}$ state, see our discussion in Sec.~\ref{sec5}.   

For a similar discussion of the relation between excitation energies 
as they follow from the RGM and the linear spin-wave theory 
for the kagome HAFM, 
see Ref.~\cite{Mueller2018}.
The ground-state excitation velocity $v/S$ 
corresponding to the linear expansion of the lowest branch $\omega_{4\mathbf{q}}$ around the $\Gamma$ point 
is shown in the inset of the lower panel of Fig.~\ref{fig03}.
Similar as for the kagome HAFM \cite{Mueller2018}, $v/S$ decreases with growing $S$.
Note that in the next section we consider the temperature dependence of the excitation energies for the PHAF, see Fig.~\ref{fig12}.

\begin{figure}
\centering 
\includegraphics[clip=on,width=80mm,angle=0]{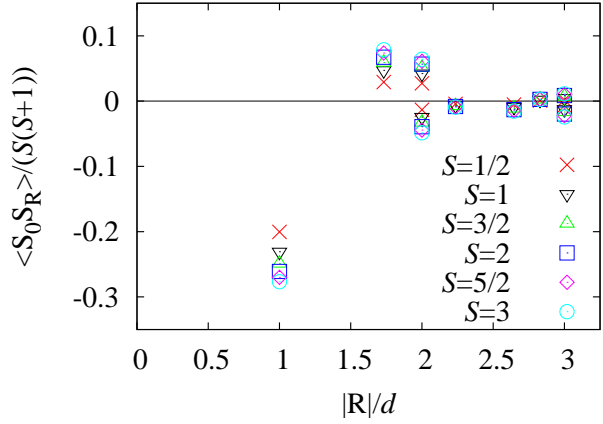}
\protect
\caption
{Ground-state correlation functions
$\langle\hat{\bm{S}}_{\boldsymbol{0}} \cdot \hat{\bm{S}}_\mathbf{R}\rangle/(S(S+1))$ 
within a range of separation $|\mathbf{R}|/d \le 3$, 
where $d=\sqrt{2}/4\approx 0.35$ is the nearest-neighbor separation,
$c_{100}<0$ ($R=d=\sqrt{2}/4\approx 0.35$),
$c_{110}>0$ ($R=\sqrt{6}/4 \approx 0.61$),
$c_{200}>0$ ($R=\sqrt{2}/2 \approx 0.71$),
$c_{2{\textrm -}20}<0$ ($R=\sqrt{2}/2 \approx 0.71$),
$c_{21{\textrm -}1}<0$ ($R=\sqrt{10}/4 \approx 0.79$),
$c_{210}<0$ ($R=\sqrt{14}/4 \approx 0.94$),
$c_{22{\textrm -}2}>0$ ($R= 1$),
$c_{300}<0$ ($R=3\sqrt{2}/4 \approx 1.06$),
and
$c_{3{\textrm -}30}>0$ ($R=3\sqrt{2}/4 \approx 1.06$),
for the quantum PHAF obtained within the minimal-version RGM 
for 
$S=1/2$ (crosses), 
$S=1$ (down-triangles), 
$S=3/2$ (up-triangles),
$S=2$ (squares),
$S=5/2$ (diamonds),
and 
$S=3$ (circles).
Note that for several separations $R$ inequivalent correlators exist.}
\label{fig04} 
\end{figure}

\begin{figure}
\centering 
\includegraphics[clip=on,width=80mm,angle=0]{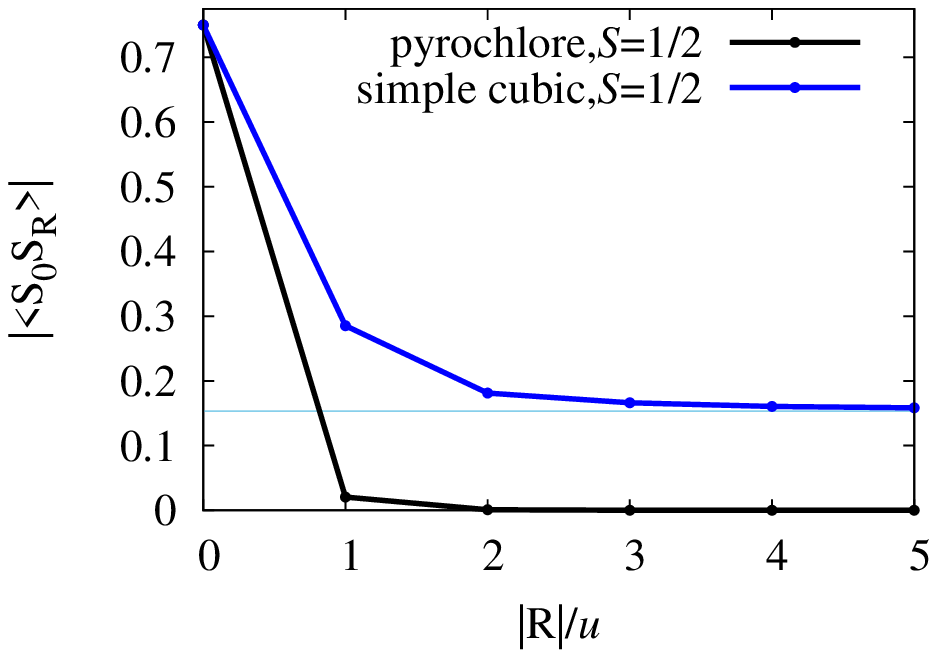}\\
\vspace{3mm}
\includegraphics[clip=on,width=80mm,angle=0]{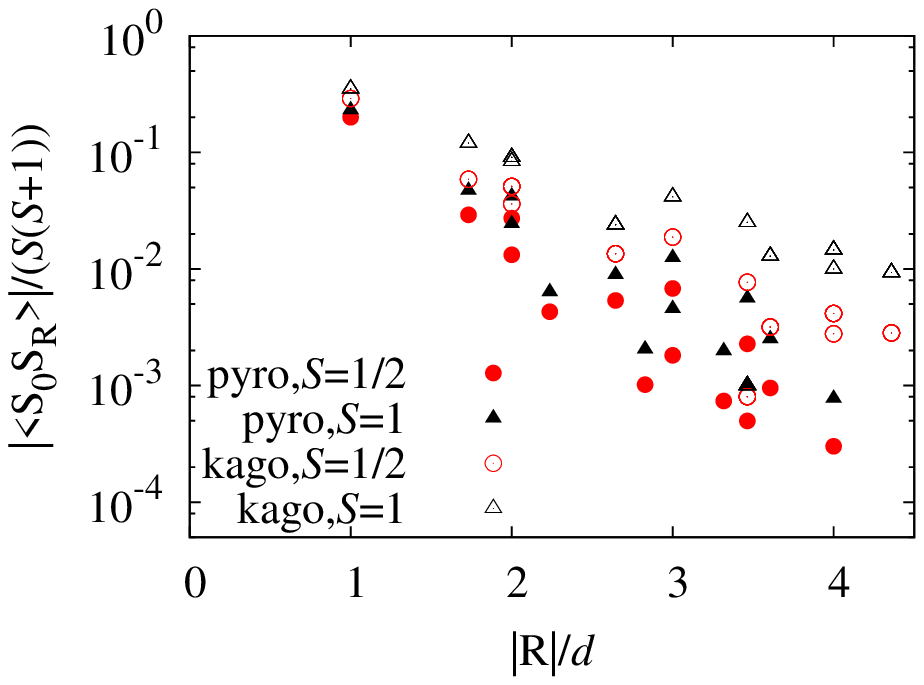}
\protect
\caption
{(Top)
The absolute value of the ground-state correlation functions $|\langle\hat{\bm{S}}_{\boldsymbol{0}}\cdot\hat{\bm{S}}_{\mathbf{R}}\rangle|$
as a function of the scaled distance $|\mathbf{R}|/u$ 
along the direction $(0,1/2,1/2)$ for the $S=1/2$ PHAF ($u=1/\sqrt{2}$, black)
and for the $S=1/2$ simple-cubic HAFM along the direction $(1,0,0)$ ($u=1$, blue).
(Bottom)
The absolute value of the ground-state correlation functions
$|\langle\hat{\bm{S}}_{\boldsymbol{0}}\cdot\hat{\bm{S}}_{\mathbf{R}}\rangle|/(S(S+1))$
as a function of the separation $|\mathbf{R}|$ (scaled by the nearest-neighbor separation $d$)
for the pyrochlore lattice ($d=\sqrt{2}/4 \approx 0.35$, filled symbols)
and
for the kagome lattice \cite{Mueller2018} ($d=1$, open symbols)
for $S=1/2$ (red) and $S=1$ (black).}
\label{fig05} 
\end{figure}

Let us turn to the spin-spin correlation functions.
In Fig.~\ref{fig04} we show all non-equivalent ground-state correlators 
$\langle\hat{\bm{S}}_{\boldsymbol{0}} \cdot \hat{\bm{S}}_\mathbf{R}\rangle/(S(S+1))$ 
up to a separation $R = |\mathbf{R}|= 3d$ 
for $S=1/2,1,\ldots,3$. 
(We use here the scaling factor $S(S+1)$ 
because it leads to an $S$-independent ground-state correlator for the isolated spin dimer with antiferromagnetic coupling.)
Since for a certain separation $R$ inequivalent sites exist, more than one data point can appear at one and the same separation $R$
(e.g., for the third-neighbor separation $R=\sqrt{2}/2$ there are two kinds of correlators, which have different signs). 
Note that the signs of the correlators coincide with the results of Ref.~\cite{Canals2000} (see Table~I in that paper).
The fast decay of the correlation functions is obvious and it is also demonstrated in Fig.~\ref{fig05}, 
where we compare the PHAF with the corresponding unfrustrated HAFM 
on the simple-cubic lattice (top) as well with the two-dimensional kagome HAFM \cite{Mueller2018} 
for spin quantum numbers $S=1/2$ and $S=1$ 
(bottom; note the logarithmic scale of the $y$-axis). 
The comparison with the simple-cubic lattice demonstrates 
the existence of a finite condensation term $C_{\mathbf{Q=(\pi,\pi,\pi)}}$ for this lattice
as well as the lack of long-range order for the PHAF.
These data may suggest an exponential decay.
Interestingly, our data also suggest that the decay of the correlation functions is faster for the PHAF.
To estimate the correlation length for the PHAF we assume such an exponential decay.
Then, a correlation length $\xi$ can be extracted using the ansatz  
$|\langle\hat{\bm{S}}_{\boldsymbol{0}}\cdot\hat{\bm{S}}_{\mathbf{R}}\rangle|\propto\textrm{exp}(-|\mathbf{R}|/\xi)$,
see Ref.~\cite{Canals1998}.
Further, we fix the direction of ${\mathbf{R}}$ to ${\bf u}=(0,1/2,1/2)$, 
i.e., ${\mathbf{R}}=n{\bf u}$, 
to have only one correlator for each separation $R=|\mathbf{R}|$, 
and consider the correlators until $n=12$.     
Using the fitting function  
$f(R)=a \,\textrm{exp}(-{R}/b)+c$
we get 
$b=0.1963 (\pm 0.09\%)$,
i.e., $\xi(T=0,S=1/2)\approx 0.20$. 
The increase of the quantum spin number $S$ leads to a slight increase of 
$|\langle\hat{\bm{S}}_{\boldsymbol{0}}\cdot\hat{\bm{S}}_{\mathbf{R}}\rangle|/(S(S+1))$
(cf. Figs.~\ref{fig04} and \ref{fig05}, bottom).
For $S=1$ we find 
$b=0.2233 (\pm 0.10\%)$, 
i.e., $\xi(T=0,S=1)\approx 0.22$,
and for $S=3$ we find
$b=0.2578 (\pm 0.38\%)$, 
i.e., $\xi(T=0,S=3)\approx 0.26$.
(Note that the fitting constant $c$ is always smaller than $10^{-4}$.)
Thus, the correlation length is less than the nearest-neighbor separation and it is even smaller than for the kagome HAFM \cite{Mueller2018}
(see also Fig.~\ref{fig05}, bottom).

\begin{figure}
\centering 
\includegraphics[clip=on,width=42.5mm,angle=0]{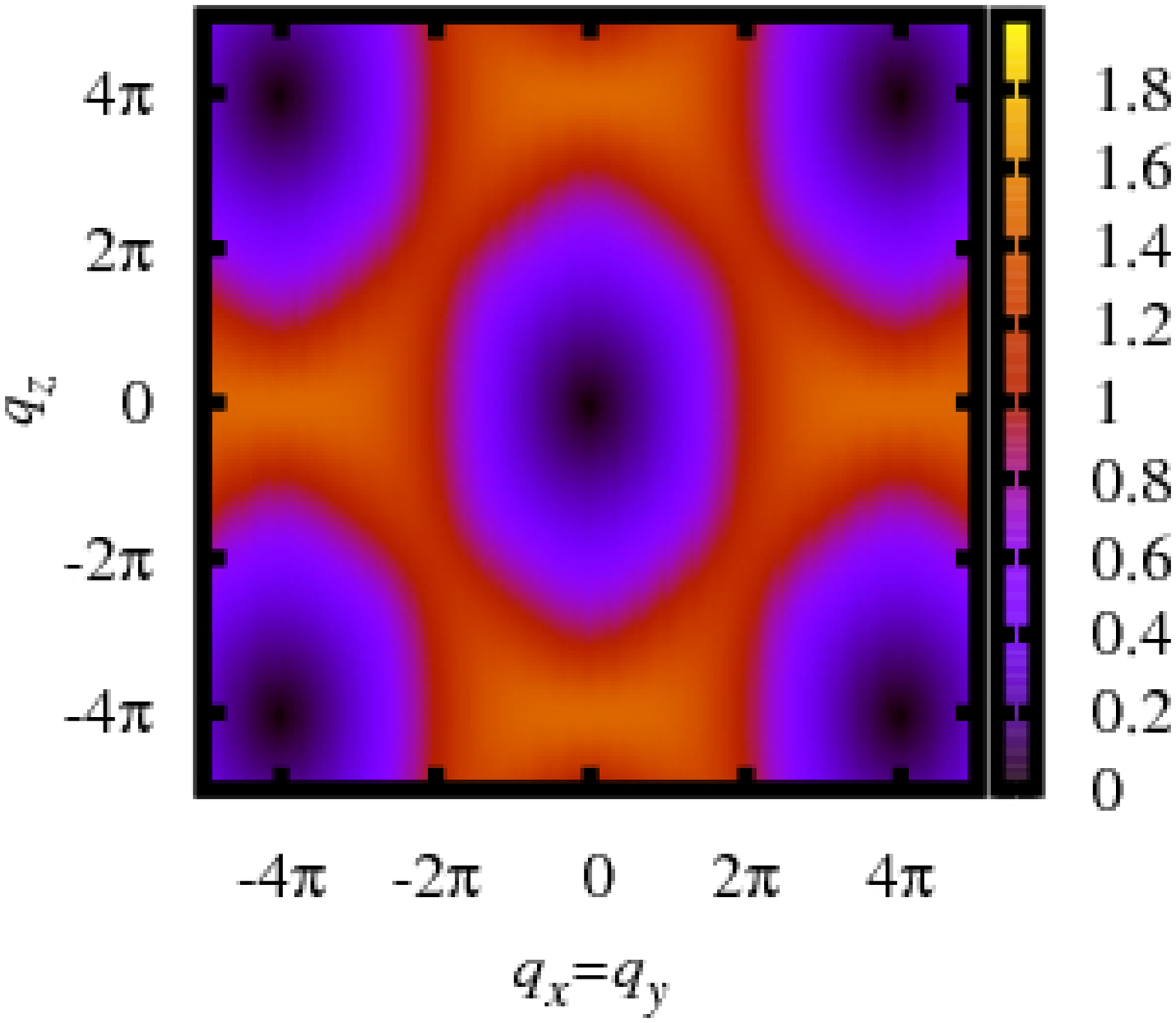}
\includegraphics[clip=on,width=42.5mm,angle=0]{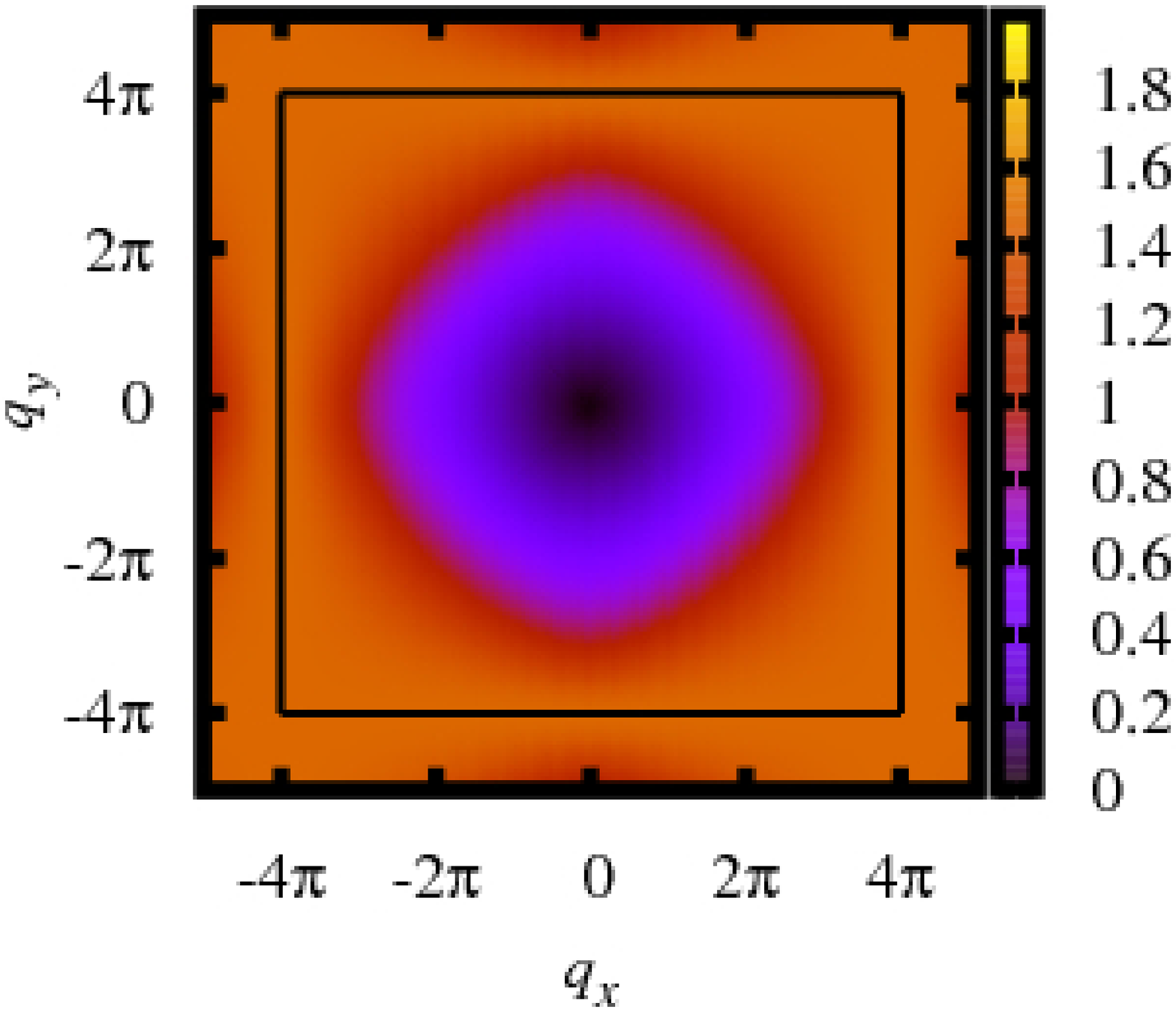}\\
\vspace{3mm}
\includegraphics[clip=on,width=42.5mm,angle=0]{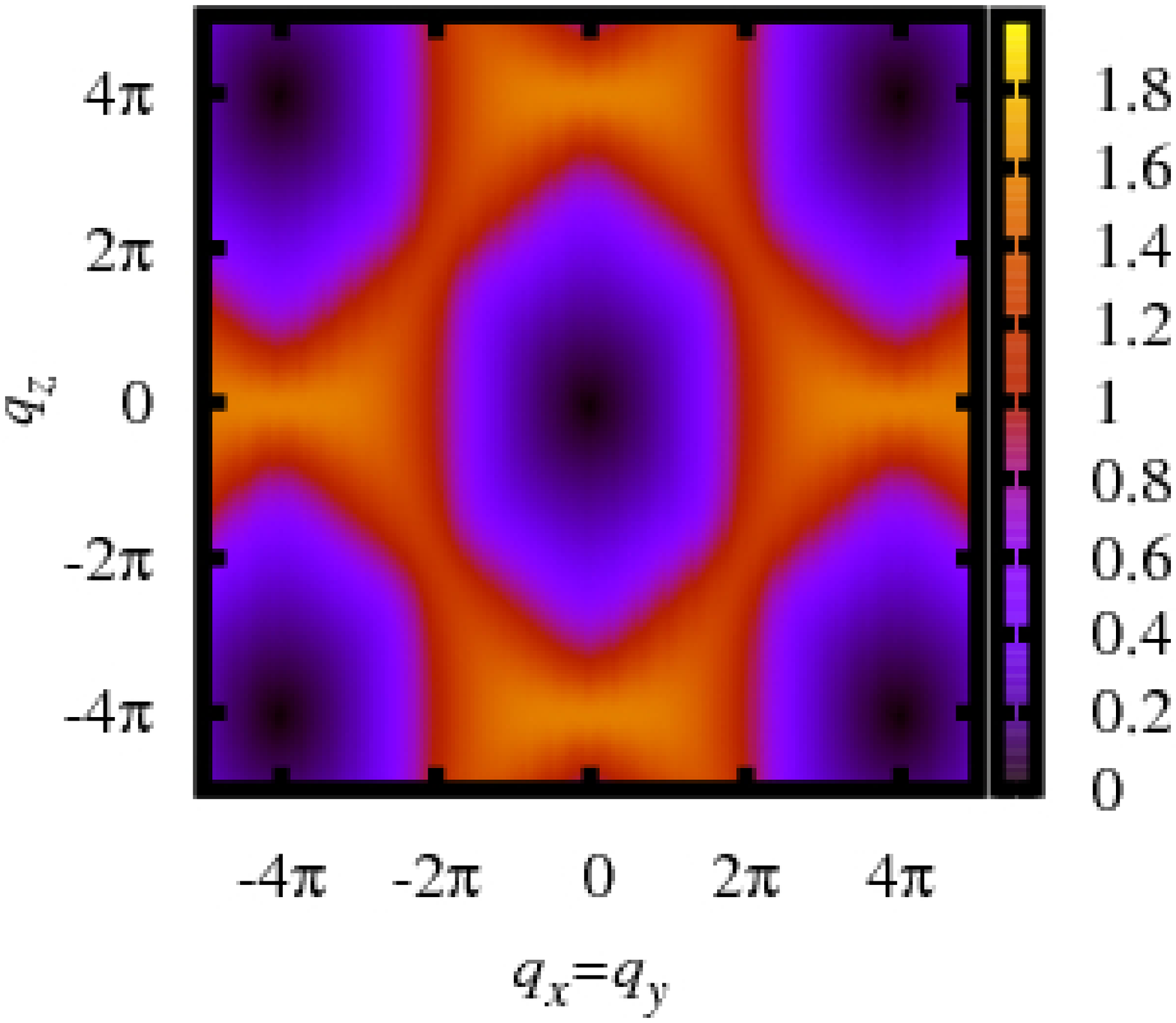}
\includegraphics[clip=on,width=42.5mm,angle=0]{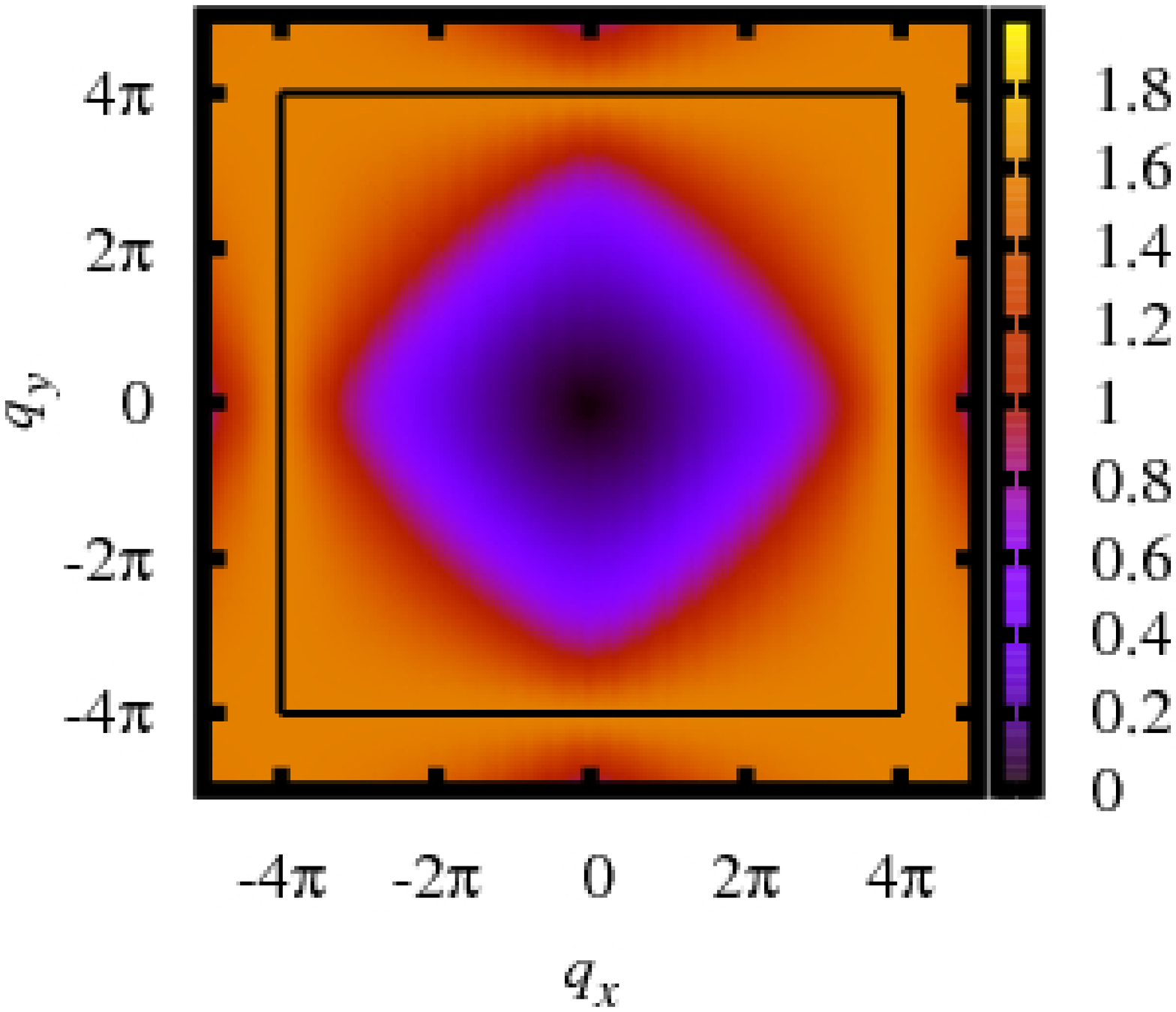}
\protect
\caption
{Normalized static structure factor $S_{\mathbf{q}}/(S(S+1))$ of the PHAF obtained within the RGM 
at $T=0$ for $S=1/2$ (top) and $S=1$ (bottom). 
We consider the two planes $q_{y}=q_{x}$ (left) and $q_{z}=0$ (right) within the (extended) Brillouin zone.
The black squares in the right panels show the ${\bf{q}}$-points which yield the (same) maximal value of $S_{(q_x,q_y,0)}$.}
\label{fig06} 
\end{figure}

\begin{figure}
\centering 
\includegraphics[clip=on,width=80mm,angle=0]{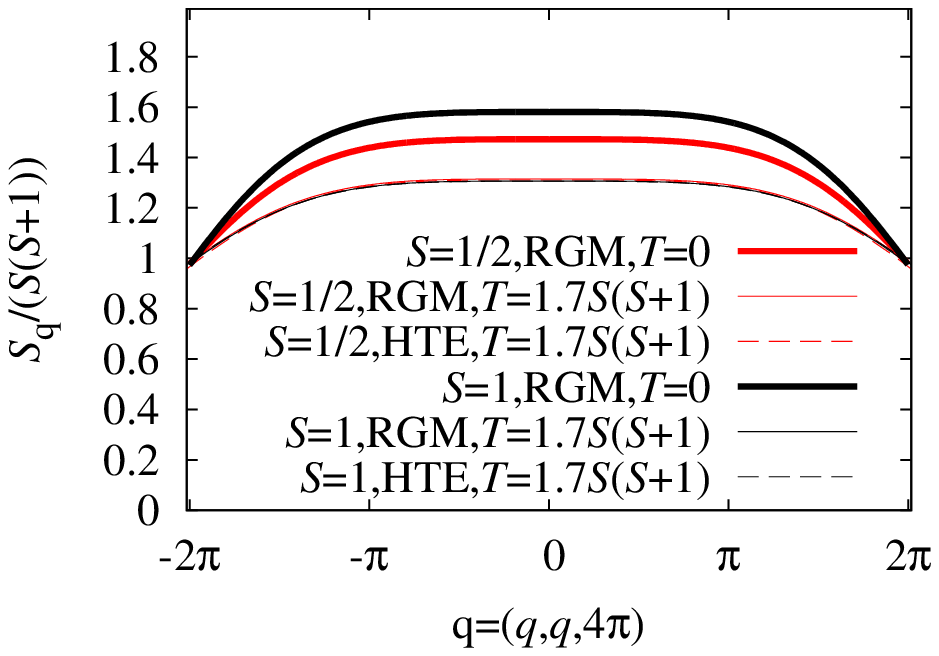}\\
\vspace{3mm}
\includegraphics[clip=on,width=80mm,angle=0]{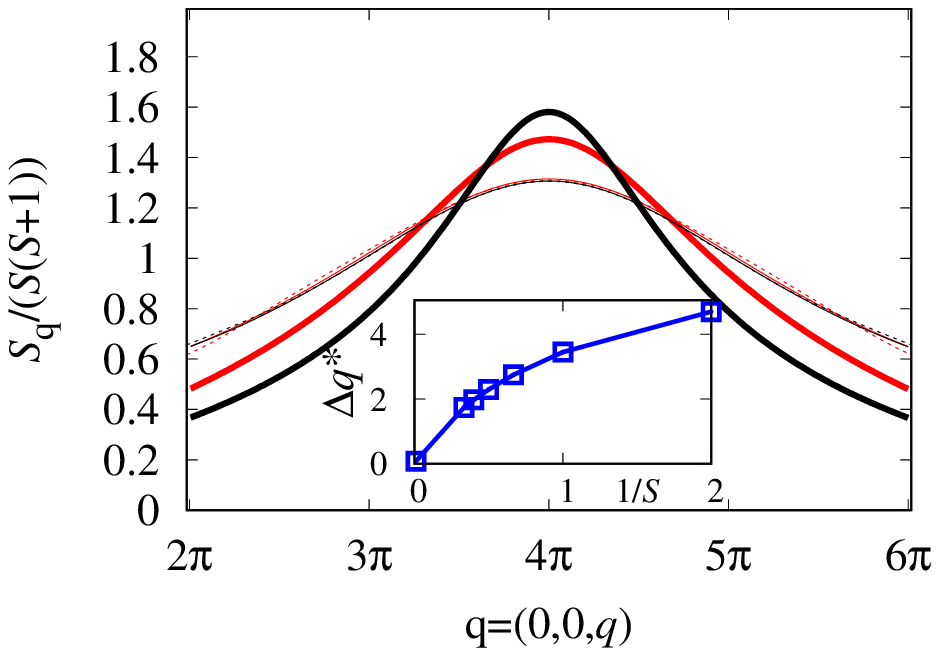}
\protect
\caption
{Horizontal cut $(q,q,4\pi)$ (top) and vertical cut $(0,0,q)$ (bottom) through the pinch point at $(0,0,4\pi)$ 
for $S=1/2$ (red) and $S=1$ (black).
RGM results at $T=0$ are shown by thick lines.
We also show the results at $T=1.7S(S+1)$ by thin lines 
(solid lines correspond to RGM results and dashed lines correspond to HTE results), 
see also Sec.~\ref{sec5}.
Note that all thin lines for $T=1.7S(S+1)$ almost coincide.
The inset (bottom) shows RGM results for the width-at-half-maximum of the pinch point $\Delta q^*$ as a function of $1/S$ at $T=0$.}
\label{fig07} 
\end{figure}

An intensity plot of the static structure factor $S_{\mathbf{q}}$, 
see, e.g., Eq.~(\ref{311}), 
is shown in Fig.~\ref{fig06} within two planes in the $\mathbf{q}$-space, 
namely,   
the $(q_{x}=q_{y})-q_z$ plane (left column)
and 
in the plane $q_{x}-q_{y}$ for $q_{z}=0$ (right column).
$S_{\mathbf{q}}$ exhibits some typical features related to spin-liquid ground states, 
which are partially also present for the kagome HAFM.
It is worth mentioning, that similar features can be seen in experiments on $S=1$ PHAF compound NaCaNi$_2$F$_7$,
see Fig.~1 and the left quadrants of Fig.~4c of Ref.~\cite{Plumb2017}.
To compare with measured data,
we notice that the neutron momentum transfer denoted in Ref.~\cite{Plumb2017} as $(h,k,l)$ 
corresponds to $(q_x/(2\pi),q_y/(2\pi),q_z/(2\pi))$
and thus, e.g., the vector $(0,0,2)$ of Ref.~\cite{Plumb2017} is the vector $(0,0,4\pi)$ in our notations.
The pinch points at, e.g., $\mathbf{Q}_{0}=(4\pi,4\pi,0)$ and other symmetry related points such as $(0,0,4\pi)$ 
indicate that each tetrahedron has vanishing total magnetization (ice rule).
Along a continuous line (within the $q_{z}=0$ -- $q_x-q_y$ plane) indicated by the black squares, 
see the right panels of Fig.~\ref{fig06},
the structure factor $S_{\mathbf{q}}/(S(S+1))$ is maximal, 
which also means that (within the numerical precision of our RGM data) $S_{\mathbf{q}}/(S(S+1))$ is constant on this line.
This remains true for the RGM data at $T>0$, see Fig.~\ref{fig16}. 
Obviously, the pinch points are located on this line of maximal $S_{\mathbf{q}}$.

For a quantitative analysis of the pinch points 
we show in Fig.~\ref{fig07} the structure factor along a horizontal and a vertical momentum cut through the pinch point at $(0,0,4\pi)$.
Since the pinch points are still present at finite temperatures 
we show in Fig.~\ref{fig07}, in addition to $T=0$, also RGM and HTE data at $T=1.7S(S+1)$.
At $T=0$ the difference between $S=1/2$ and $S=1$ is noticeable, 
but there is practically no difference between the two cases at $T=1.7S(S+1)$. 
Moreover, the agreement between RGM and HTE data at this temperature is very good. 
Along the horizontal cut, $S_{\bf{q}}/(S(S+1))$ remains almost constant in a pretty wide region of $q$-values.
Along the vertical cut across the pinch point 
the sharpening of $S_{\bf{q}}/(S(S+1))$ as $S$ increases from $1/2$ to $1$ is obvious,
see the thick red and black lines in Fig.~\ref{fig07}, bottom (see also Fig.~\ref{fig16} in Sec.~\ref{sec5}).  
To quantify this sharpening,
we plot in the inset in Fig.~\ref{fig07} (bottom) the width of the pinch point at the half of the maximum $\Delta q^*$ as a function of $1/S$ at $T=0$.
Note that in the classical limit
the pinch points become sharper as $\sqrt{T}$ as $T$ decreases, see Ref.~\cite{Zhang2018}.
The sharpening of the pinch points is related to the decreasing role of quantum fluctuation as $S$ increases. 
Only in the classical limit each tetrahedron can have vanishing total spin, 
whereas perfect spin-singlet formation on all tetrahedra is not possible in the quantum model, 
since the total spin of a tetrahedron does not commute with the Hamiltonian.
Note that similar features were observed in Ref.~\cite{FPRG_Pyro_2018}.

\begin{figure}
\centering 
\includegraphics[clip=on,width=80mm,angle=0]{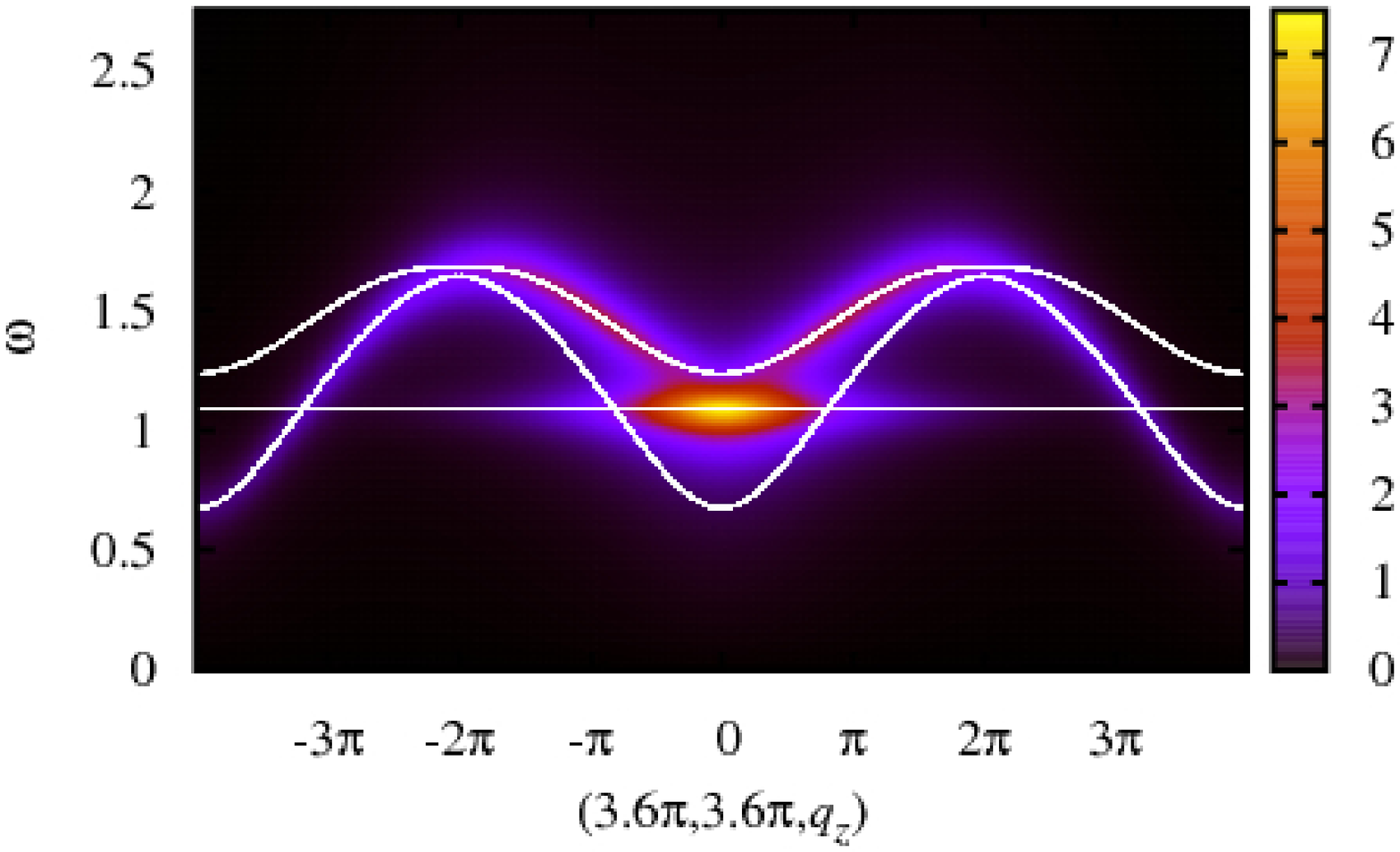}\\
\vspace{3mm}
\includegraphics[clip=on,width=80mm,angle=0]{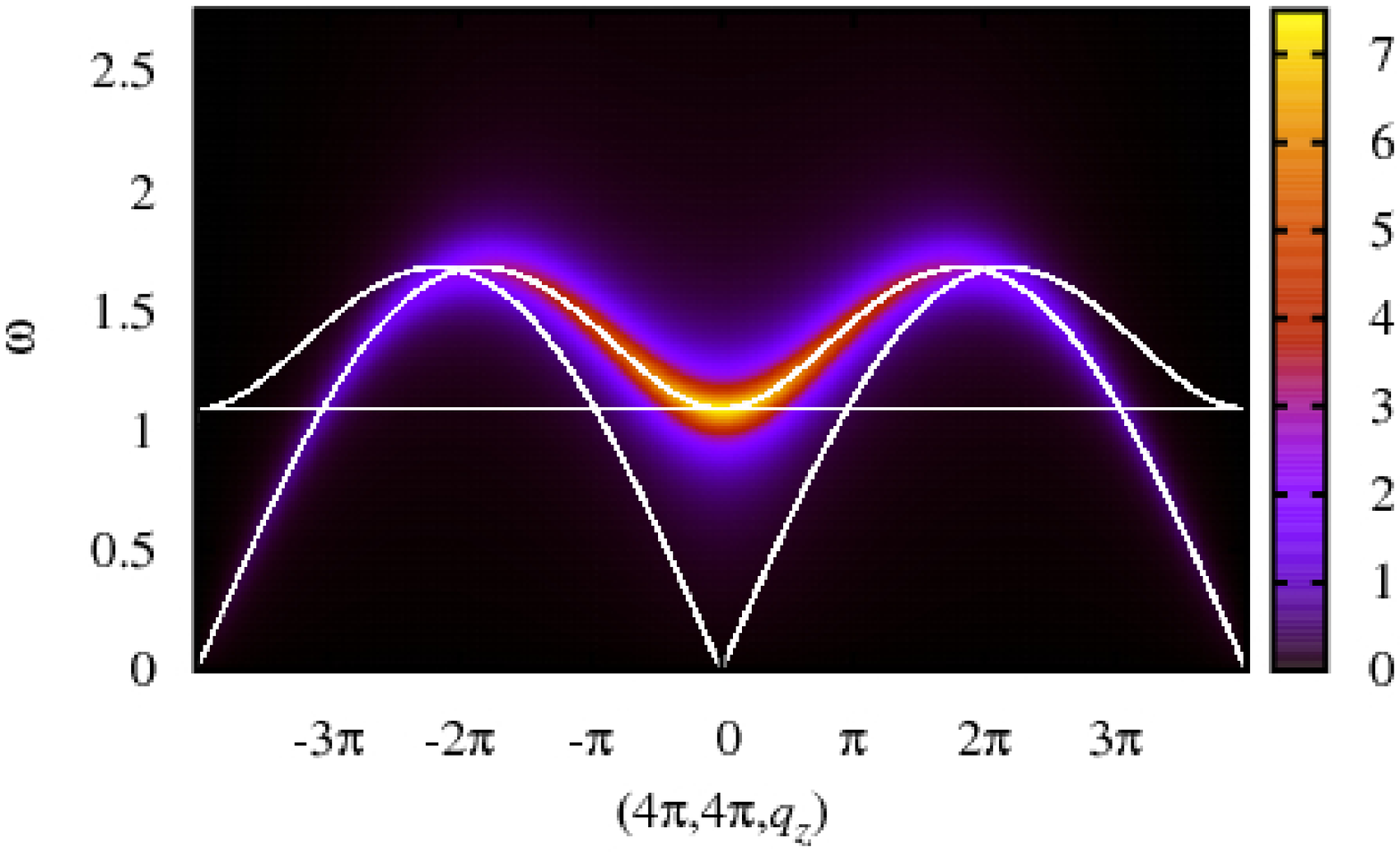}\\
\vspace{3mm}
\includegraphics[clip=on,width=80mm,angle=0]{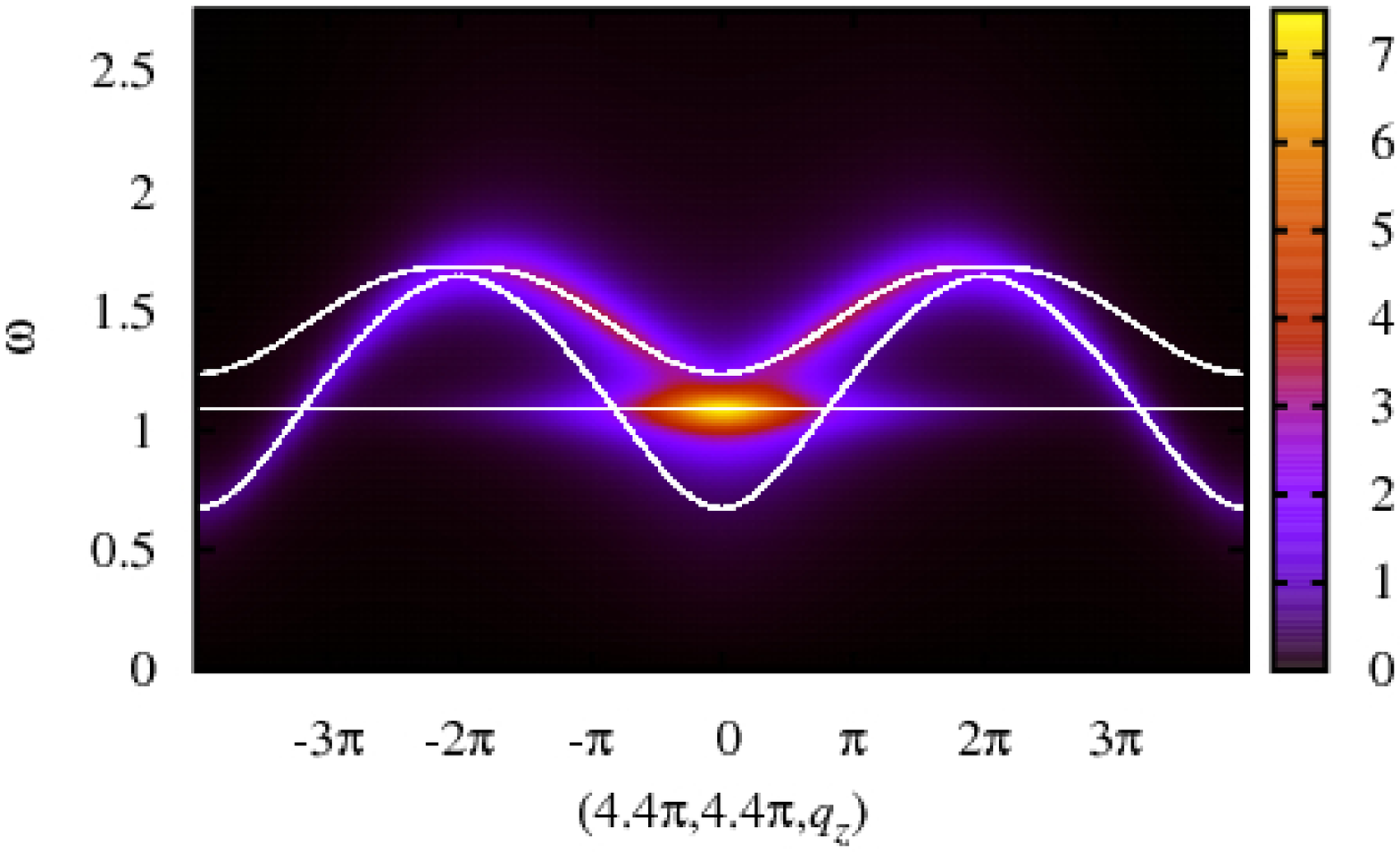}
\protect
\caption
{Dynamical structure factor $S^{zz}_{\mathbf{q}}(\omega)$ of the $S=1/2$ PHAF ($J=1$)
at $q_x=q_y=3.6\pi,\,4\pi, 4.4\pi$ along the line $-4\pi\le q_z\le 4\pi$ for $T=0$. 
We set $\epsilon=0.1$.
The white lines correspond to the excitation energies $\omega_{\gamma{\mathbf{q}}}$ (\ref{303}).}
\label{fig08} 
\end{figure}

\begin{figure}
\centering 
\includegraphics[clip=on,width=80mm,angle=0]{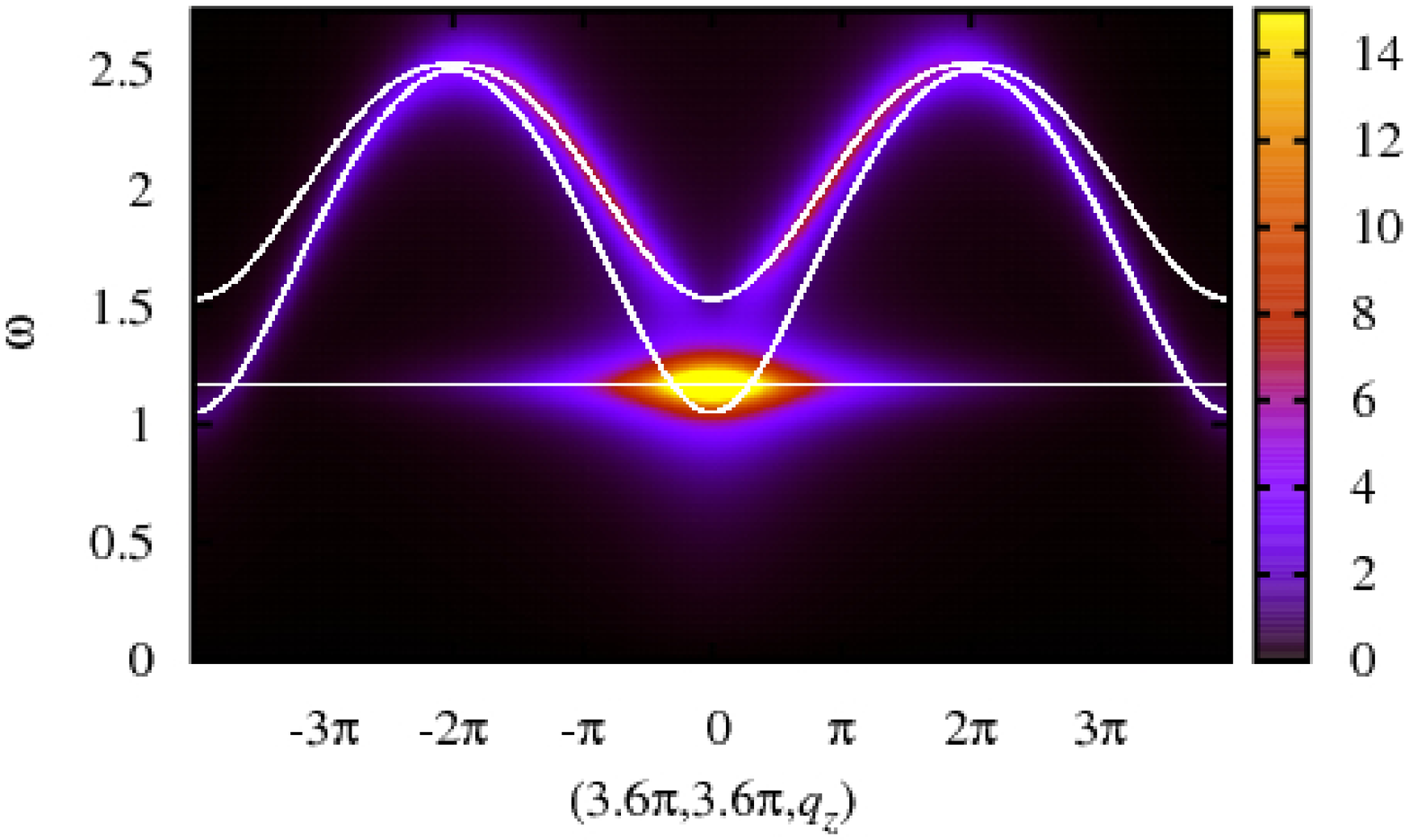}\\
\vspace{3mm}
\includegraphics[clip=on,width=80mm,angle=0]{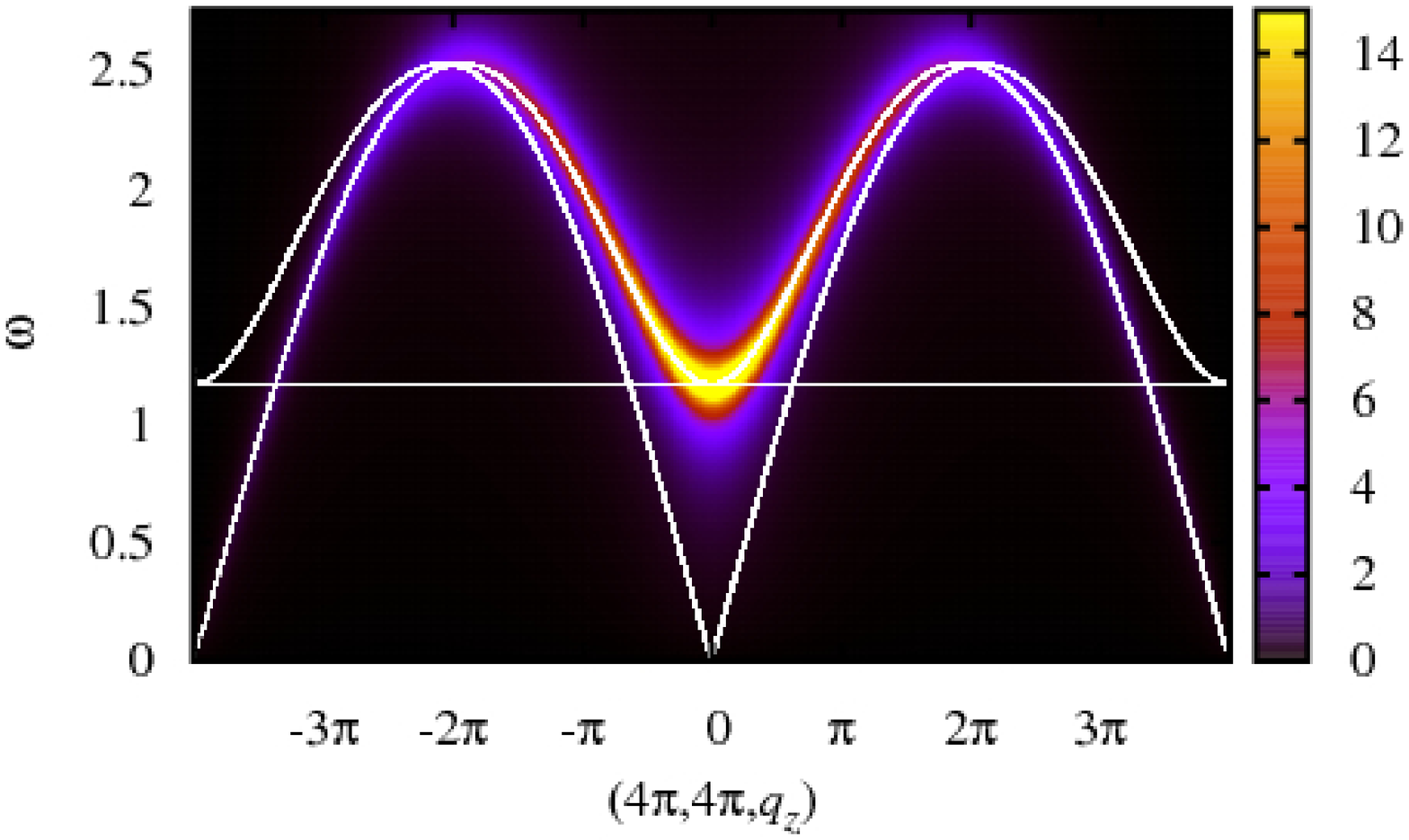}\\
\vspace{3mm}
\includegraphics[clip=on,width=80mm,angle=0]{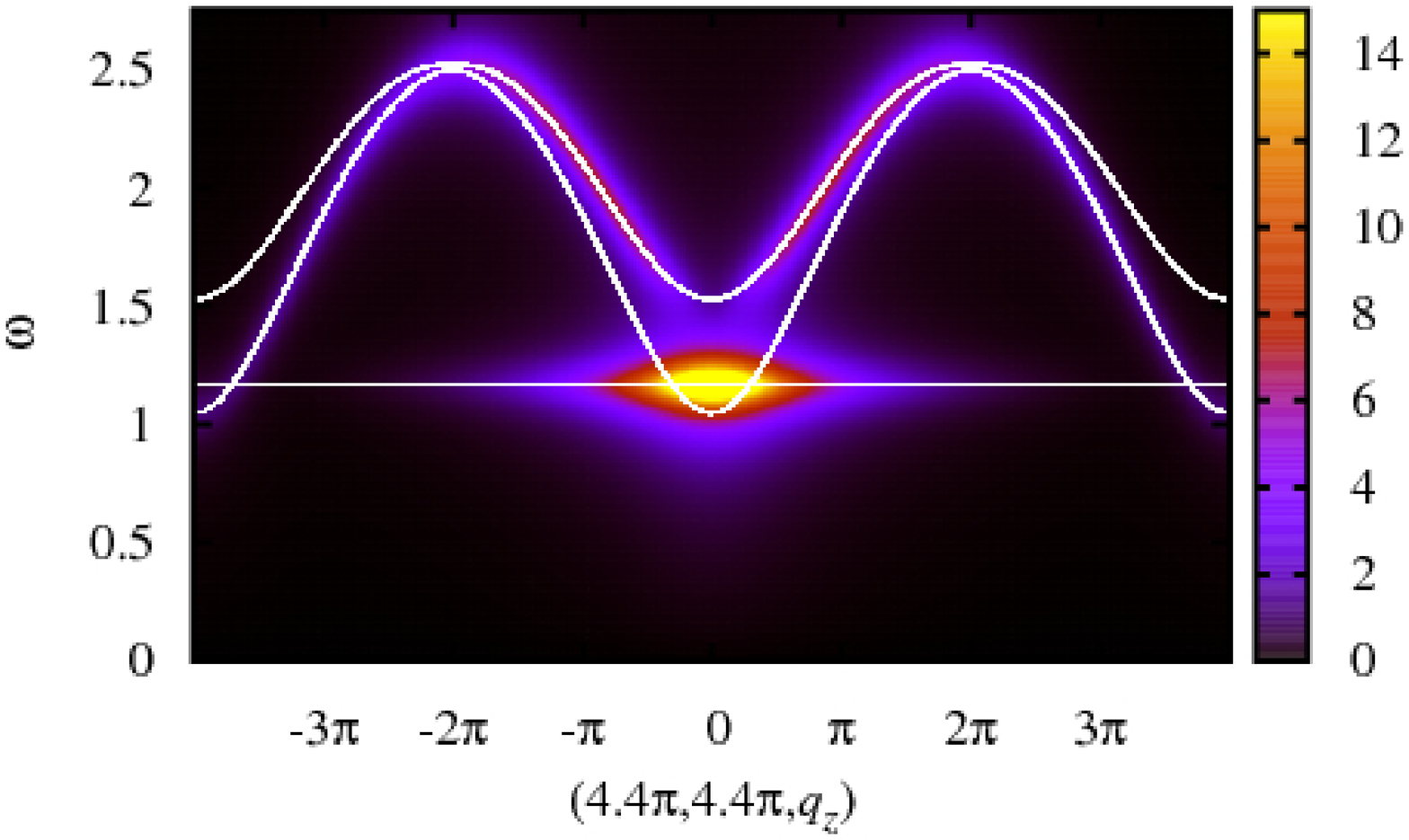}
\protect
\caption
{Dynamical structure factor $S^{zz}_{\mathbf{q}}(\omega)$ of the $S=1$ PHAF ($J=1$)
at $q_x=q_y=3.6\pi,\,4\pi, 4.4\pi$ along the line $-4\pi\le q_z\le 4\pi$ for $T=0$. 
We set $\epsilon=0.1$.
The white lines correspond to the excitation energies $\omega_{\gamma{\mathbf{q}}}$ (\ref{303}).
These theoretical plots may be compared to experimental data reported in the left part of Fig.~3a of Ref.~\cite{Plumb2017}.}
\label{fig09} 
\end{figure}

\begin{figure}
\centering 
\includegraphics[clip=on,width=80mm,angle=0]{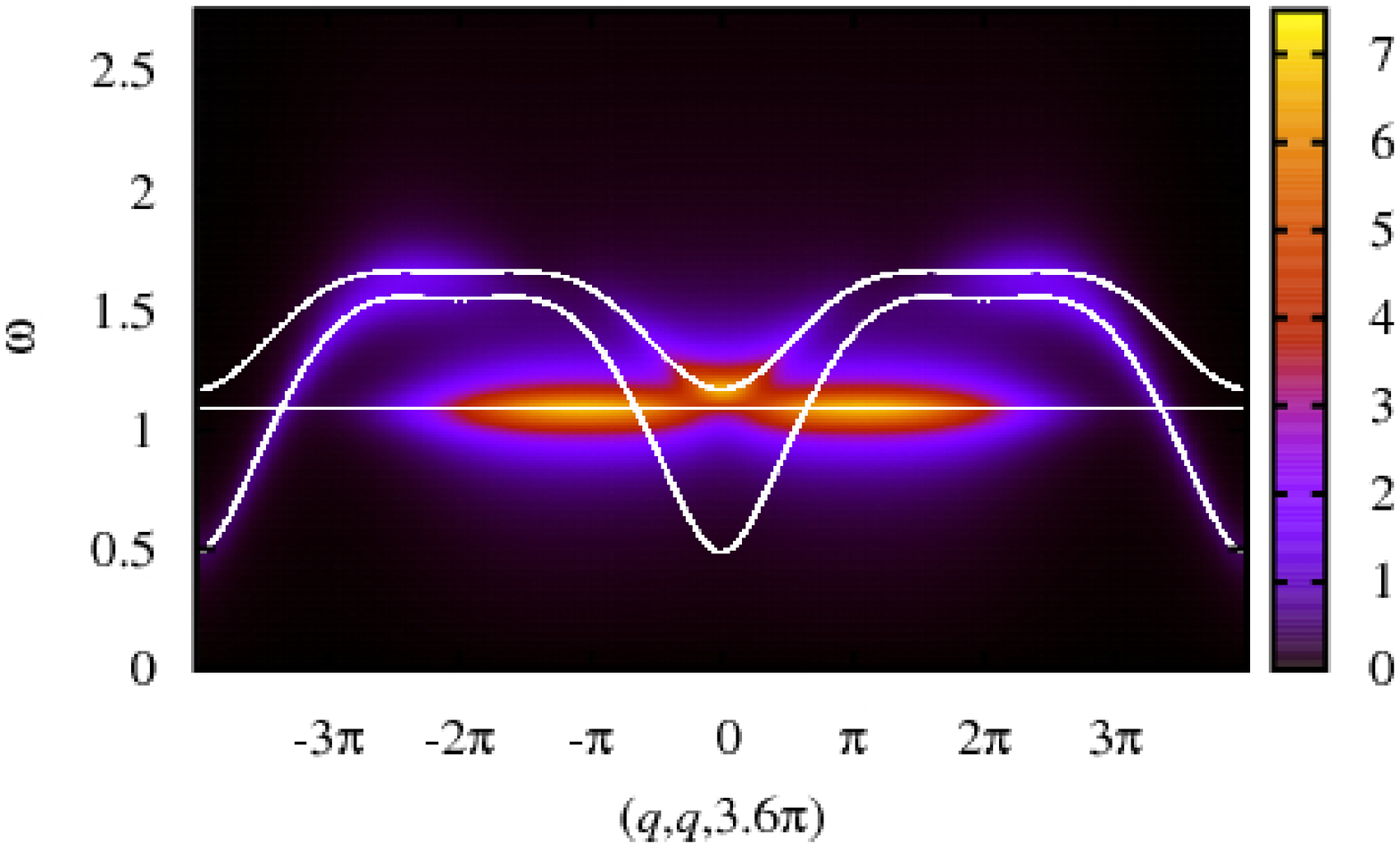}\\
\vspace{3mm}
\includegraphics[clip=on,width=80mm,angle=0]{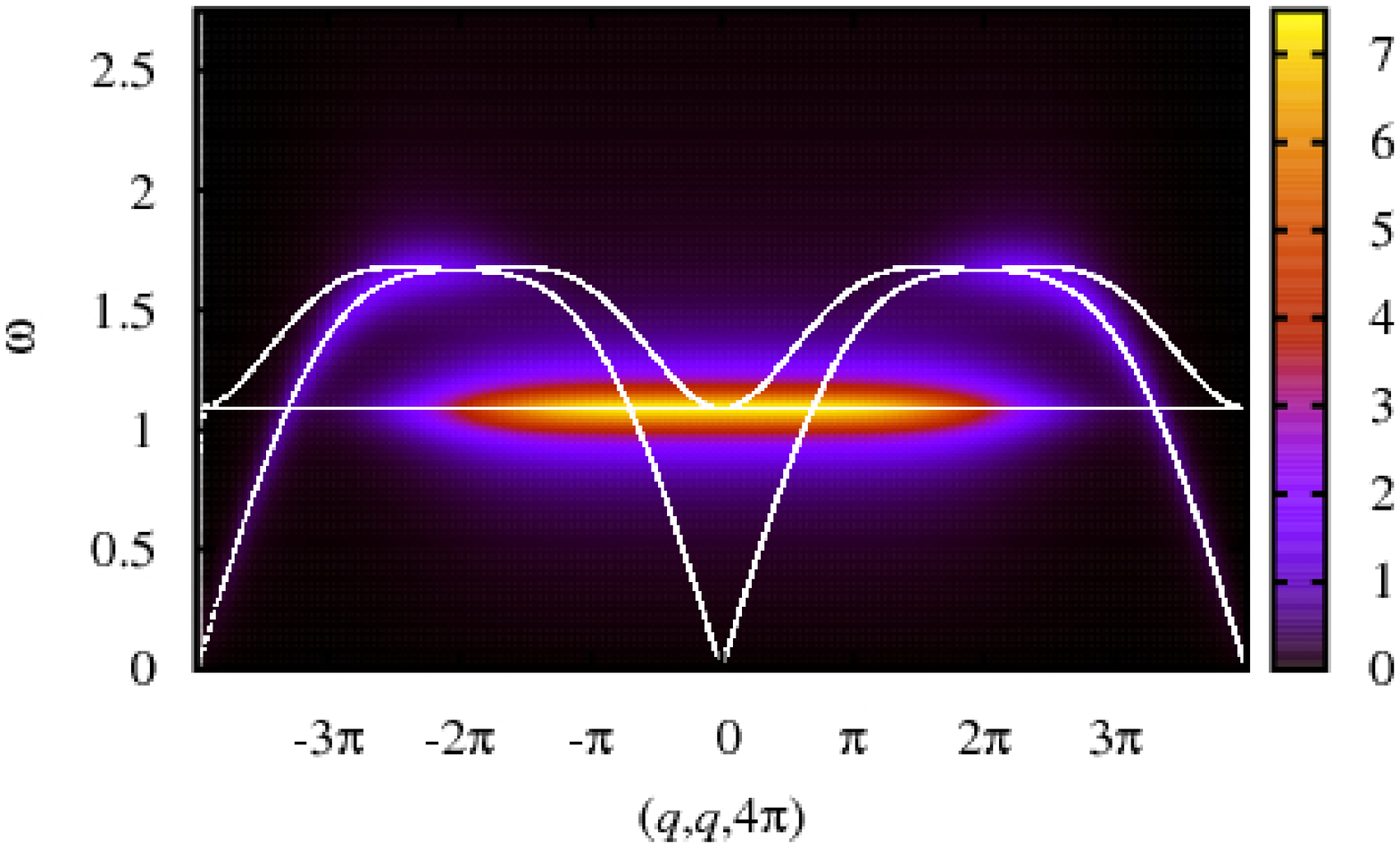}\\
\vspace{3mm}
\includegraphics[clip=on,width=80mm,angle=0]{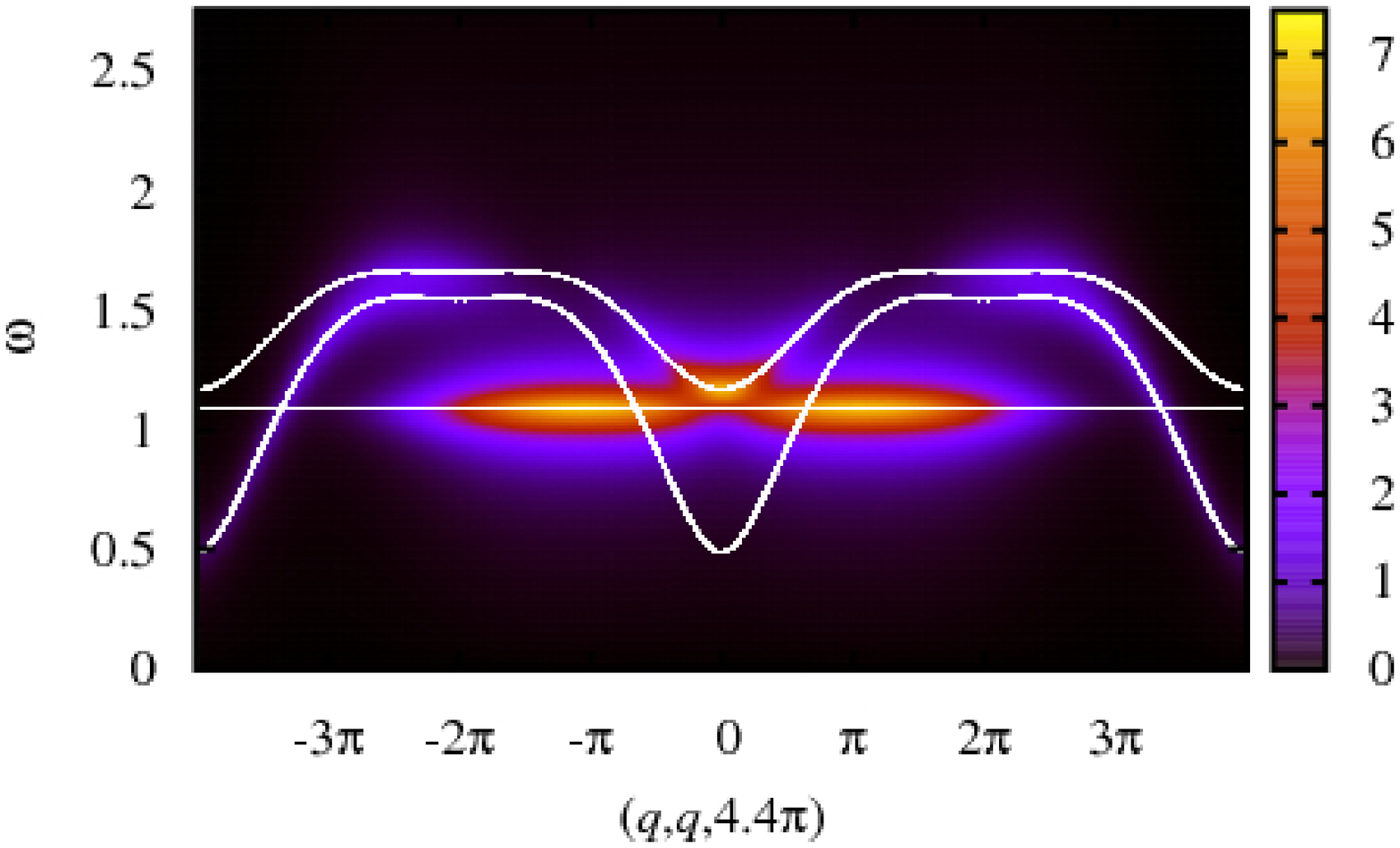}
\protect
\caption
{Dynamical structure factor $S^{zz}_{\mathbf{q}}(\omega)$ of the $S=1/2$ PHAF ($J=1$)
at $q_z=3.6\pi,\,4\pi, 4.4\pi$ along the line $-4\pi\le q_x=q_y\le 4\pi$ for $T=0$. 
We set $\epsilon=0.1$.
The white lines correspond to the excitation energies $\omega_{\gamma{\mathbf{q}}}$ (\ref{303}).}
\label{fig10} 
\end{figure}

\begin{figure}
\centering 
\includegraphics[clip=on,width=80mm,angle=0]{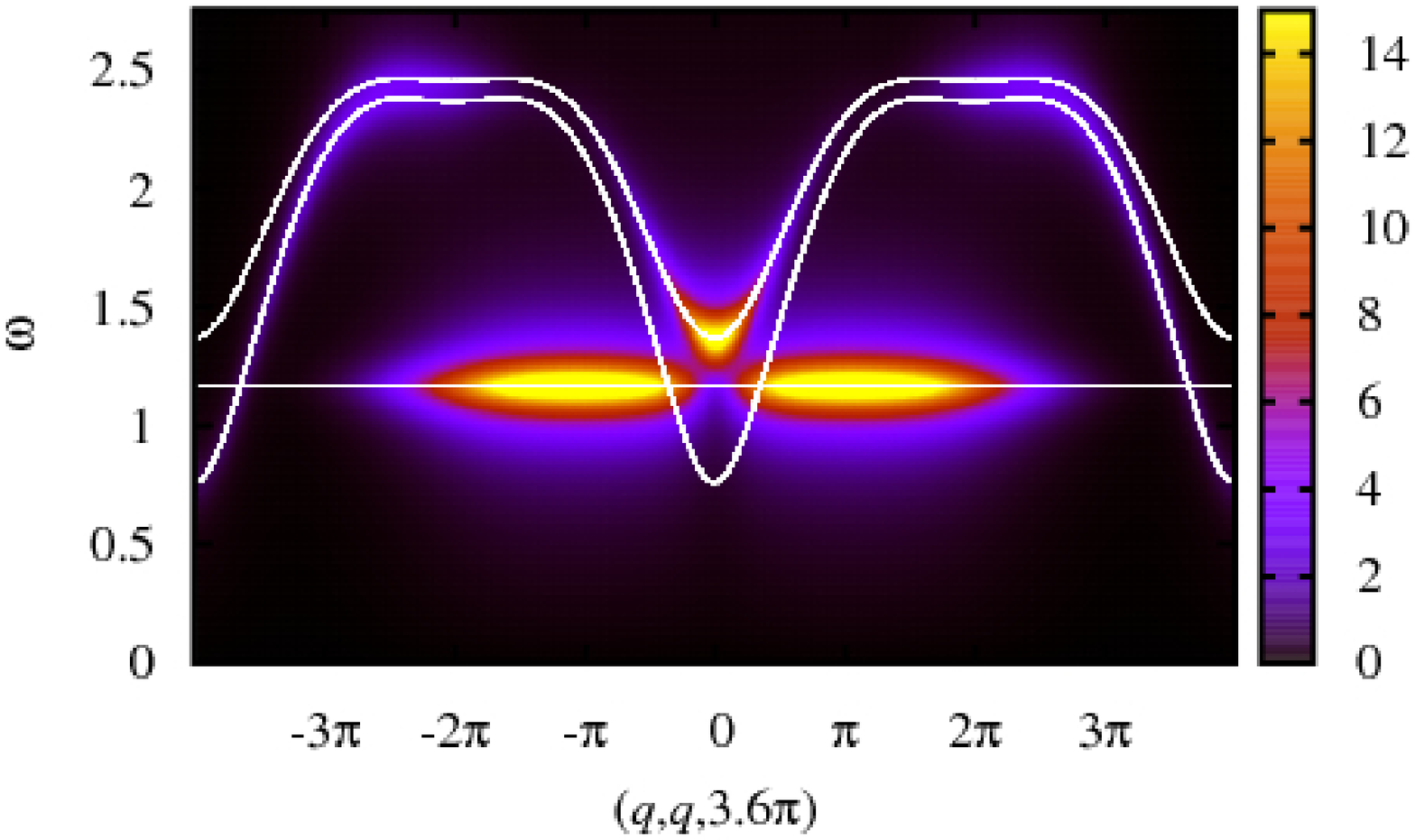}\\
\vspace{3mm}
\includegraphics[clip=on,width=80mm,angle=0]{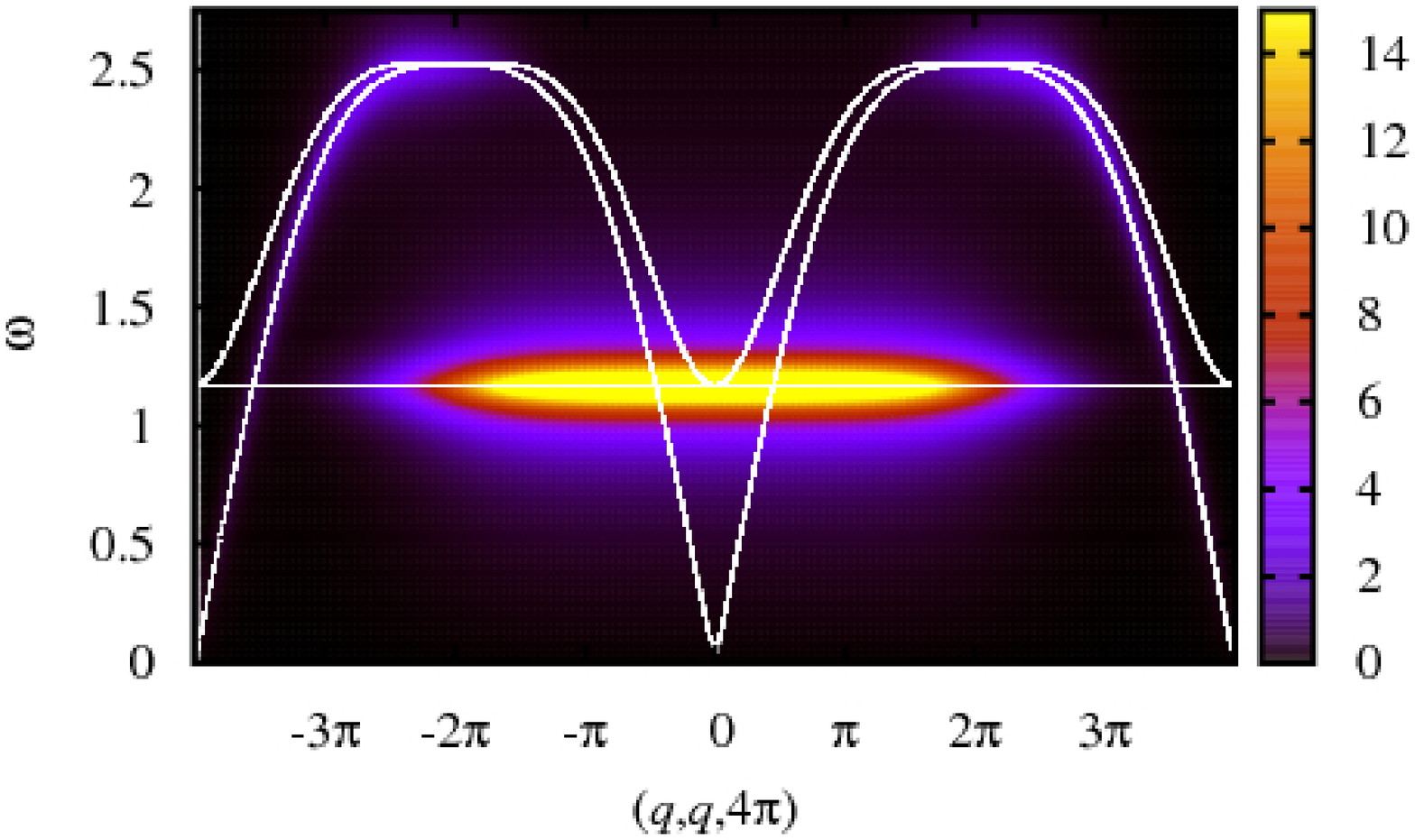}\\
\vspace{3mm}
\includegraphics[clip=on,width=80mm,angle=0]{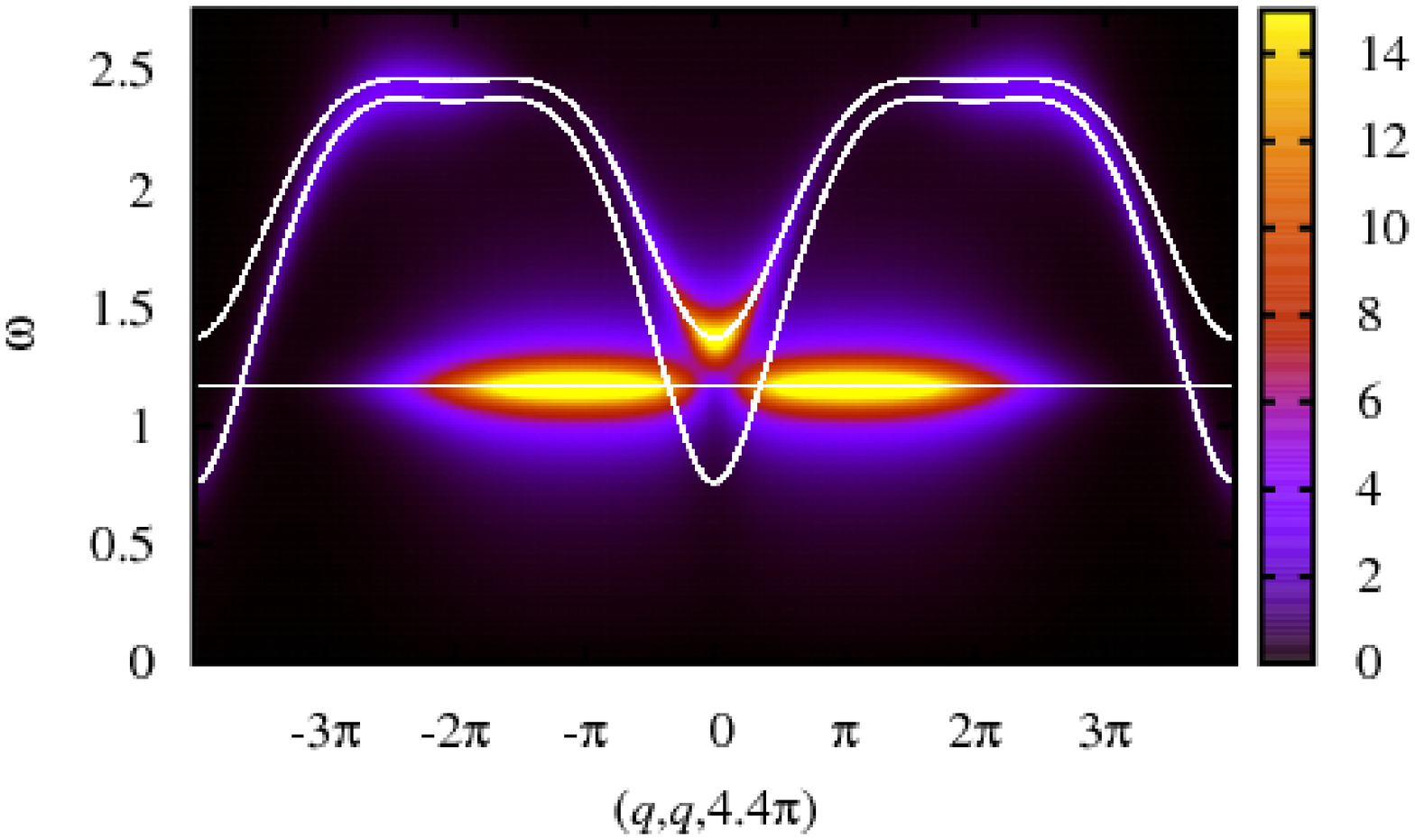}
\protect
\caption
{Dynamical structure factor $S^{zz}_{\mathbf{q}}(\omega)$ of the $S=1$ PHAF ($J=1$)
at $q_z=3.6\pi,\,4\pi, 4.4\pi$ along the line $-4\pi\le q_x=q_y\le 4\pi$ for $T=0$. 
We set $\epsilon=0.1$.
The white lines correspond to the excitation energies $\omega_{\gamma{\mathbf{q}}}$ (\ref{303}).
Our theoretical plots may be compared to experimental data reported in the right part of Fig.~3a of Ref.~\cite{Plumb2017}.}
\label{fig11} 
\end{figure}

Next we consider the dynamical structure factor $S_{{\bf{q}}}^{zz}(\omega)$, see Eq. (\ref{308}).
While dynamical quantities for the quantum HAFM on the kagome lattice were discussed in several theoretical papers,
see, e.g., Refs.~\cite{Sherman2018,Halimeh2018,Yan2018},
corresponding results for the quantum PHAF are scarce.
Very recently
a combination of molecular dynamics simulations, stochastical dynamical theory and linear spin-wave theory
has been used for a theoretical study of the dynamical structure factor of the spin-1 pyrochlore material NaCaNi$_2$F$_7$ \cite{Zhang2018}.
Corresponding experimental data for the dynamical properties of NaCaNi$_2$F$_7$ can be found in Ref.~\cite{Plumb2017}. 
Here we also use the experimental data shown in Figs.~2 and 3 of Ref.~\cite{Plumb2017} as a guideline
for the presentation of our RGM results for $S_{{\bf{q}}}^{zz}(\omega)$ given in Figs.~\ref{fig08}, \ref{fig09}, \ref{fig10}, and \ref{fig11}.
To connect our calculations to this compound, 
we recall that for NaCaNi$_2$F$_7$ the estimate for $J$ is about 3.2~meV (37~K).
Then the experiments at $T=1.5$~K correspond to $T/J\approx 0.04$ in our study
and the energy transfers 2~meV, 8~meV, and 12~meV correspond to $\omega/J\approx 0.625$, 2.5, and 3.75, respectively.  
We also recall that the neutron momentum transfer denoted in Ref.~\cite{Plumb2017} as $(h,k,l)$ 
corresponds to $(q_x/(2\pi),q_y/(2\pi),q_z/(2\pi))$ in our notations. 

We begin with the momentum cut along $(4\pi,4\pi,q_z)$, 
see Figs.~\ref{fig08} (for $S=1/2$) and \ref{fig09} (for $S=1$) and the corresponding Fig.~3a of Ref.~\cite{Plumb2017}.
Except for the low-frequency region,
our theoretical predictions look similar to the experimental observations,
both having a kind of vertical fountain structure with the origin at ${\bf{q}}=(4\pi,4\pi,q_z=0)$ and $\omega\approx J$.
The nonzero values of $S^{zz}_{\bf{q}}(\omega)$ at ${\bf{q}}=(4\pi,4\pi,q_z)$ 
shown in the middle panels of Fig.~\ref{fig08} ($S=1/2$) and Fig.~\ref{fig09} ($S=1$)
are (nonuniformly) concentrated only along the dispersive branch $\omega_{3{\bf{q}}}$ with $\vert q_z\vert\le 2\pi$.
Since experiments give the scattering cross-section at $(q,q,q_z)$ with $3.6\pi<q<4.4\pi$,
we show in Figs.~\ref{fig08} and \ref{fig09} theoretical predictions for $q=3.6\pi$ (upper panels) and $q=4.4\pi$ (lower panels), too.
These slight deviations from $q_x=q_y=4\pi$ change the scattering dramatically.
Namely,
the dynamical structure factor is now concentrated mostly along the dispersionless excitation branch $\omega_{1{\bf{q}}}=\omega_{2{\bf{q}}}$ around $\vert q_z\vert\le\pi$.
Although the dispersive branch $\omega_{3{\bf{q}}}$ is still visible, 
the value of $S^{zz}_{\bf{q}}(\omega)$ along this branch is relatively small.
The comparison of the cases $S=1/2$ and $S=1$ does not show qualitative differences,
however, all features for the latter case are much sharper.

For another momentum cut, 
${\bf{q}}=(q,q,4\pi)$,
see Figs.~\ref{fig10} and \ref{fig11} and the corresponding Fig.~3a of Ref.~\cite{Plumb2017},
$S^{zz}_{\bf{q}}(\omega)$ is again concentrated along one excitation branch,
but now along the dispersionless one $\omega_{1{\bf{q}}}=\omega_{2{\bf{q}}}$ with $\vert q\vert\le 2\pi$.
When $q_z$ deviates from $4\pi$ (in experiments $3.6\pi<q_z<4.4\pi$),
$S^{zz}_{\bf{q}}(\omega)$ redistributes in the $q-\omega$ plane, 
i.e., it vanishes along the dispersionless branch around $q=0$, 
but emerges for these $q$-values along the dispersive branch $\omega_{3{\bf{q}}}$.
This looks similar to what can be seen in the experimental data around ${\bf{q}}=(q,q,4\pi)$, $\vert q\vert\le 2\pi$ and $\omega\approx J$,
cf. the right part of Fig.~3a of Ref.~\cite{Plumb2017}.
Again, all features become sharper as $S$ increases from 1/2 to 1.

To conclude this discussion,
apparently, the RGM results can reproduce the experimentally observed features at 3 \ldots 8~meV 
shown in the left and right parts of Fig.~3a of Ref.~\cite{Plumb2017}
(see Fig.~\ref{fig09} and Fig.~\ref{fig11}, respectively),
but not the $\bf q$-independent features below 2~meV.
We mention that a similar disagreement at low frequencies between theory and experiment was reported in Ref.~\cite{Zhang2018}.
A possible origin of this discrepancy may consist in disorder 
(there is Na$^{1+}$/Ca$^{2+}$ charge disorder which is expected to generate a random variation in the magnetic exchange interactions)
and/or further small terms in the Hamiltonian 
(the nearest-neighbor $3\times 3$ exchange interaction matrix has three more components
the values of which are, however, smaller than $0.1$~meV)
relevant for the specific magnetic compound studied in the experiment, 
see the corresponding discussion in Ref.~\cite{Zhang2018}.   
Let us finally mention that the experimental data for the dynamical structure factor
are obtained at a finite temperature $T=1.8$~K \cite{Plumb2017}. 
Bearing in mind the exchange constant $J \approx 3.2$~meV (37~K) of NaCaNi$_2$F$_7$,  
we have $T/(S(S+1)) \approx 0.025 J$, which practically corresponds to zero temperature, 
see Fig.~\ref{fig12}, where the temperature dependence of the excitation spectrum is shown.

\section{Finite-temperature properties} 
\label{sec5}
\setcounter{equation}{0}

In this section we consider only the extreme quantum cases $S=1/2$ and $S=1$
and report below the RGM results along with the HTE results.
As mentioned in Sec.~\ref{sec4} and discussed in Ref.~\cite{Mueller2018} 
the minimal version considered here works best for low spin quantum numbers $S$.
Moreover, for the particular cases of the kagome HAFM and the PHAF for larger $S$ 
the RGM equations may lead at finite temperatures to unphysical poles in the specific heat \cite{PHD_Muller}. 
To overcome this drawback one needs an additional input to open the possibility to consider more vertex parameters \cite{Mueller2018}.     

\begin{figure}
\centering 
\includegraphics[clip=on,width=80mm,angle=0]{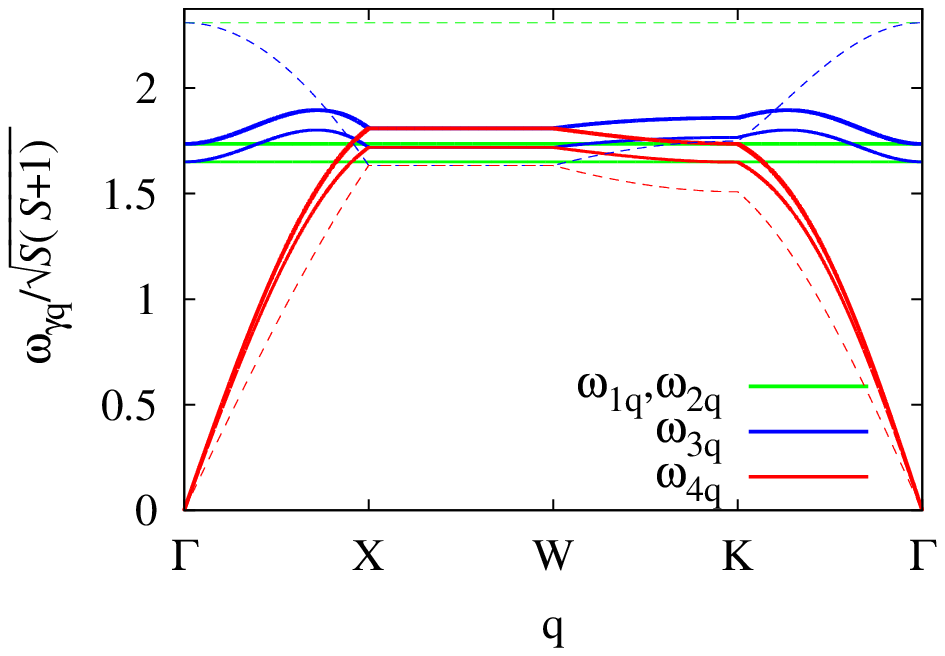}\\
\vspace{3mm}
\includegraphics[clip=on,width=80mm,angle=0]{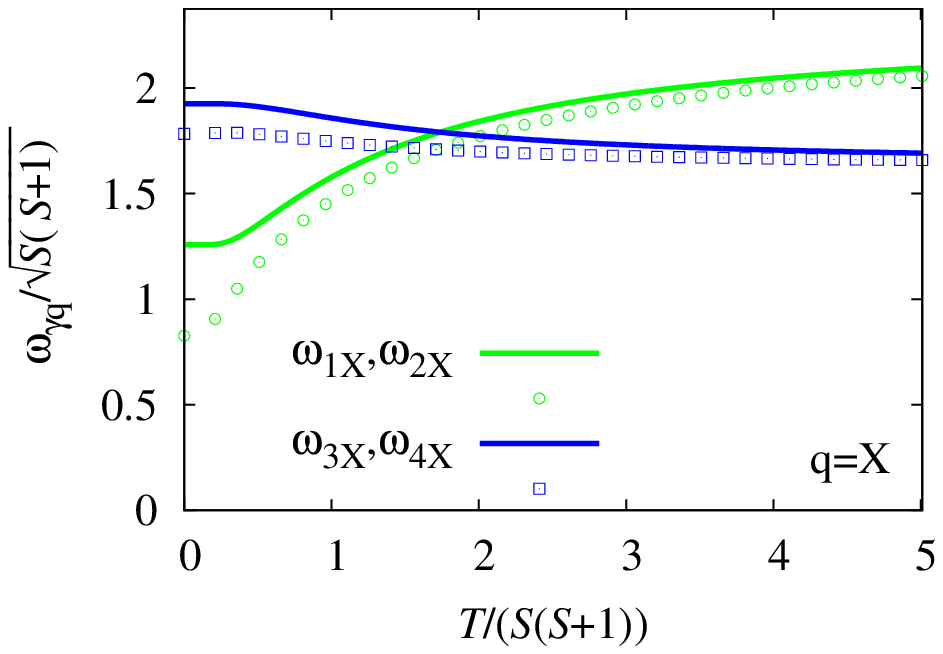}
\protect
\caption{(Top)
Dispersion of the excitation energies $\omega_{\gamma\mathbf{q}}/\sqrt{S(S+1)}$
(Eq.~(\ref{303}), $J=1$)
at temperature $T=1.5$ for $S=1/2$ (thick) and $S=1$ (thin) and in the infinite-temperature limit $T\to\infty$ (very thin dashed). 
Note that $\omega_{\gamma\mathbf{q}}/\sqrt{S(S+1)}$ is independent of $S$ at $T\to\infty$.
The points $\Gamma$, X, W, and K in the first Brillouin zone of a face-centered-cubic Bravais lattice 
are given by $\Gamma=(0,0,0)$, X$=(0,2\pi,0)$, W$=(\pi,2\pi,0)$, and K$=(3\pi/2,3\pi/2,0)$, 
see Fig.~\ref{fig01}, bottom. 
(Bottom)
Temperature dependence of the excitation energies $\omega_{\gamma\mathbf{q}}/\sqrt{S(S+1)}$ at the X point for $S=1/2$ (lines) and $S=1$ (symbols).
Note that at the X point $\omega_{3\mathbf{q}}=\omega_{4\mathbf{q}}$ is valid for all temperatures.}
\label{fig12} 
\end{figure}

We start with the discussion of the RGM results for the excitations.
As mentioned above the RGM provides an improved description of the excitation spectrum compared to linear spin-wave theory. 
Since the excitation energies $\omega_{\gamma{\bf{q}}}$ contain spin correlation functions $c_{ijk}$, cf. Eq.~(\ref{303}), 
they are temperature dependent.
At $T\to \infty$, 
we have $c_{ijk}=0$ resulting in the simplified expressions
$\omega_{1{\bf{q}}}^2/J^2=\omega_{2{\bf{q}}}^2/J^2=16S(S+1)/3$,
$\omega_{3{\bf{q}}}^2/J^2=4S(S+1)(2+D_{\mathbf{q}})/3$,
and
$\omega_{4{\bf{q}}}^2/J^2=4S(S+1)(2-D_{\mathbf{q}})/3$.
Note that in this limit
$\omega_{\gamma{\bf{q}}}^2$ does not depend on the sign of $J$ and scales as $S(S+1)$.
The branches of the spectrum (\ref{303}) in the ground state shown in Fig.~\ref{fig03}, top,
can be compared with those shown in Fig.~\ref{fig12}, top, for $T \to \infty$ and for $T=1.5$.
The temperature dependence of the dispersive bands at the X point and of the flat bands are shown in Fig.~\ref{fig12}, bottom.  
Note that the flat-band excitations increase monotonously with $T$ and become the highest-energy ones at about $T/(S(S+1)) \approx 1.8$.

\begin{figure}
\centering 
\includegraphics[clip=on,width=80mm,angle=0]{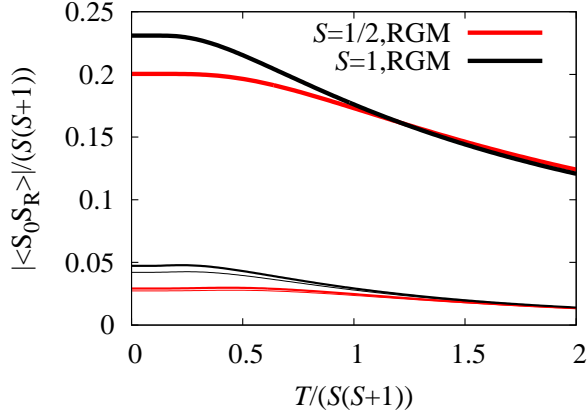}
\protect
\caption
{The absolute value of correlation functions $|\langle\hat{\bm{S}}_{\boldsymbol{0}}\cdot\hat{\bm{S}}_{\mathbf{R}}\rangle|/(S(S+1))$
between  nearest neighbors (thick), next-nearest neighbors (normal), and between third-nearest neighbors straight along two bonds (thin), 
as a function of the normalized temperature $T/(S(S+1))$ ($J=1$) 
for $S=1/2$ (red) and $S=1$ (black). 
Note that for $S=1/2$ the next-nearest- and third-nearest-neighbor correlators almost coincide.}
\label{fig13} 
\end{figure}

RGM data for the temperature dependence of the spin correlations for nearest, next-nearest and third neighbors are presented in Fig.~\ref{fig13}.
These short-range correlators show almost no dependence on $T$ at low temperatures.
For the nearest-neighbor and next-nearest-neighbor correlators 
the decrease of $|\langle\hat{\bm{S}}_{\boldsymbol{0}}\cdot\hat{\bm{S}}_{\mathbf{R}}\rangle|/(S(S+1))$
for $T/(S(S+1)) \gtrsim 0.5$ is noticeable. 
For the third-neighbor correlator (being already very small at $T=0$) the influence of $T$ is very weak.    

\begin{figure}
\centering 
\includegraphics[clip=on,width=80mm,angle=0]{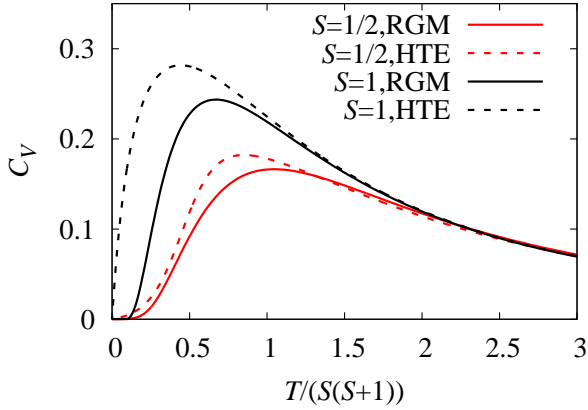}
\protect
\caption
{Specific heat of the PHAF 
obtained by RGM (solid) and HTE (dashed, Pad\'{e} [6,7] for $S=1/2$ and Pad\'{e} [5,6] for $S=1$) 
as a function of the normalized temperature $T/(S(S+1))$ ($J=1$)
for $S=1/2$ (red) and $S=1$ (black).}
\label{fig14} 
\end{figure}

Next we present in  Fig.~\ref{fig14} the RGM and the HTE results for the temperature dependence of the specific heat $C_V(T)$. 
In the high-temperature region the HTE and the RGM results coincide down to about $T/(S(S+1)) \approx 1$.
The temperature profile $C_V(T)$ is typical for spin systems with only short-range order.
The increase and the shift of the main maximum with growing $S$ known for the kagome HAFM \cite{Mueller2018} is also present for the PHAF, 
cf. also Ref.~\cite{Lohmann2014}.
At low temperatures for strongly frustrated quantum magnets unconventional features in the temperature profile of the specific heat, 
such as shoulders or additional maxima may appear, 
see, e.g., Refs.~\cite{Misguich2005,Mun:WJCMP14,Shimokawa2016,BML:npjQM18,kago42}.
We do not find such peculiar features for the PHAF within our RGM approach.
However, we do not claim that the RGM is able to detect the subtle role of low-lying excitations 
relevant for such particular low-temperature properties.

\begin{figure}
\centering 
\includegraphics[clip=on,width=80mm,angle=0]{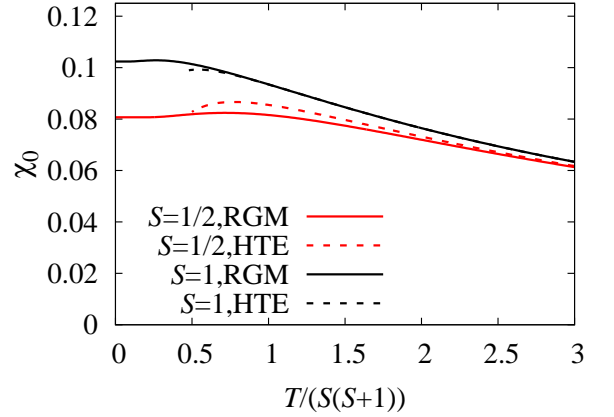}
\protect
\caption
{Uniform susceptibility of the PHAF
within the RGM (solid) and the HTE (dashed, Pad\'{e} [6,7] for $S=1/2$ and Pad\'{e} [5,6] for $S=1$) 
as a function of the normalized temperature $T/(S(S+1))$ ($J=1$)
for $S=1/2$ (red) and $S=1$ (black).}
\label{fig15} 
\end{figure}

A straightforward outcome from the RGM equations is the susceptibility $\chi_{\mathbf{Q}}$ given in Eq.~(\ref{307}).
In Fig.~\ref{fig15} we report the temperature dependence of the uniform  susceptibility $\chi_{\bf{0}}$ of the $S=1/2$ and $S=1$ PHAF
obtained within the RGM and HTE approaches. 
Again at high temperatures the results of both approaches coincide.
The temperature dependence of $\chi_{\bf{0}}$ is smooth and the typical maximum is weakly pronounced. 

\begin{figure}
\centering 
\includegraphics[clip=on,width=80mm,angle=0]{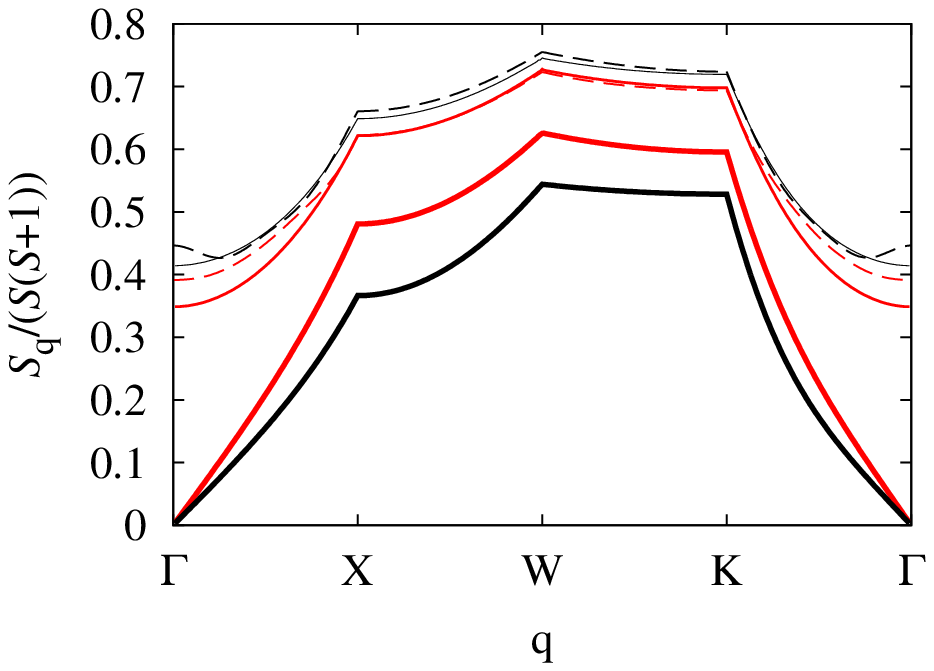}\\
\vspace{3mm}
\includegraphics[clip=on,width=80mm,angle=0]{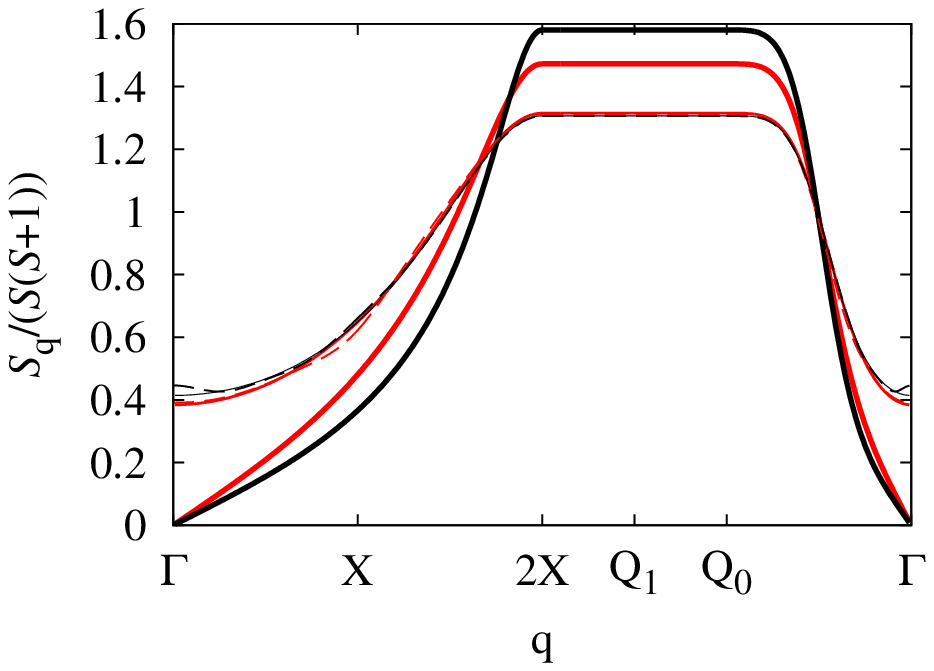}
\protect
\caption{Normalized static structure factor $S_{\mathbf{q}}/(S(S+1))$ along two paths in ${\bf {q}}$-space
for $S=1/2$ (red) and $S=1$ (black) 
obtained within RGM (solid) at $T=0$ (thick) and $T=1.7S(S+1)$ (thin) 
compared with HTE data for $T=1.7S(S+1)$ (thin dashed, 12th order for $S=1/2$ and 10th order for $S=1$).
Here $\Gamma=(0,0,0)$, X$=(0,2\pi,0)$, W$=(\pi,2\pi,0)$, and K$=(3\pi/2,3\pi/2,0)$, see Fig.~\ref{fig01}, bottom.
The points 2X$=(0,4\pi,0)$, ${\bf{Q}}_1=(2\pi,4\pi,0)$, and ${\bf{Q}}_0=(4\pi,4\pi,0)$ 
are located along the upper line of the black square in the right part of Fig.~\ref{fig06}.}
\label{fig16} 
\end{figure}

A quantity of high interest in frustrated magnets is the (static) magnetic structure factor (\ref{311})
which is related to an experimentally accessible quantity, 
the differential magnetic neutron cross section ${\rm{d}}\sigma/{\rm{d}}\Omega$.
Already in Fig.~\ref{fig06}
we have presented a contour plot of the ground-state structure factor of the PHAF in two planes of the ${\bf{q}}$-space,
namely,
$q_x=q_y$ (left panels)
and
$q_z=0$ (right panels).
Since the spin correlations are already at zero temperature extremely short-ranged,  
the influence of $T$ on $S_{\bf{q}}$ is weak and the basic features of $S_{\bf{q}}$ shown in Fig.~\ref{fig06} survive at moderate temperatures.
To get a more quantitative information on the temperature dependence of $S_{\bf{q}}$   
we compare the $\bf{q}$-dependence of the structure factor for $T=0$ and $T=1.7S(S+1)$ for $S=1/2$ and $S=1$ in Fig.~\ref{fig16}. 
Here,
the ${\bf{q}}$-line chosen for the upper panel is the same as in Figs.~\ref{fig03} and \ref{fig12},
whereas the ${\bf{q}}$-line chosen for the lower panel contains the path 
along which the structure factor reaches its maximal value $S^{\rm max}_{\bf{q}}$ 
(see the black-square line in the right panels of Fig.~\ref{fig06}).
As can be seen from Fig.~\ref{fig16}, 
the line of maximal $S_{\bf{q}}$ remains horizontal at finite $T$ 
and $S^{\rm max}_{\bf{q}}$ decreases only by 11\% (17\%) for $S=1/2$ ($S=1$) as increasing the temperature from $T=0$ to $T=1.7S(S+1)$.  
We mention that the temperature dependence of momentum cuts through a pinch point can be found in Fig.~\ref{fig07}. 

It is obvious from Fig.~\ref{fig16} 
that the static structure factors of the PHAF obtained by the RGM and the HTE are in good agreement for the selected temperature of $T=1.7S(S+1)$, 
where the 12th order HTE for $S=1/2$ is reliable in the whole Brillouin zone.
For $S=1$ we only can present data for the 10th order HTE. 
Although the overall-agreement in this case is still good, 
the HTE shows slight oscillations near the $\Gamma$ point $\mathbf{q} \approx \mathbf{0}$.
We also mention that our $S_{\bf{q}}$ data are in good agreement with recent PFFRG results, 
see Fig.~14 in Ref.~\cite{FPRG_Pyro_2018}.   

\begin{figure}
\centering 
\includegraphics[clip=on,width=80mm,angle=0]{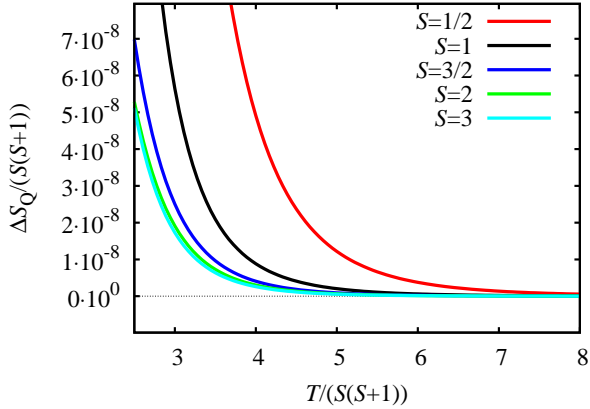}
\protect
\caption
{Difference of the static structure factor 
$\Delta S_{\mathbf{Q}}=S_{\mathbf{Q}_{1}}-S_{\mathbf{Q}_{0}}$ (scaled by $S(S+1)$)
with $\mathbf{Q}_{1}=(2\pi,4\pi,0)$ and $\mathbf{Q}_{0}=(4\pi,4\pi,0)$
within the HTE approach for $S=1/2$ (12th order) and $S=1,3/2,2,3$ (10th order) 
as a function of the normalized temperature $T/(S(S+1))$.}
\label{fig17} 
\end{figure}

As mentioned in Sec.~\ref{sec4}, 
within the numerical accuracy of the RGM 
the magnitude of the static structure factor along the black square in the right panels of Fig.~\ref{fig06} 
(line of maximal height, including the points 2X, ${\bf{Q}}_1$, and ${\bf{Q}}_0$) 
is the same, cf. also the lower panel in Fig.~\ref{fig16}.
Although the HTE treatment is restricted to high temperatures,
nevertheless it may provide rigorously some important information about the PHAF properties, 
such as {\it order by disorder} selection of magnetic structures. 
We will use the analytical HTE expressions for ${S}_{\bf{q}}$
to extract information on the behavior of the structure factor along the line of maximal height. 
We also go beyond the extreme quantum cases $S=1/2, 1$ and show results for $S>1$ for comparison. 
In Table~\ref{tab1} we present the HTE series of ${S}_{\bf q}$ up to the 9th order 
along the line ${\mathbf{q}}=(q_x,4\pi,0)$ including the points 2X, ${\bf{Q}}_1$, and ${\bf{Q}}_0$. 
(In Appendix~A, 
we provide the first three HTE terms of the PHAF static structure factor at arbitrary ${\mathbf{q}}$ points.) 
We observe that the ${\bf{q}}$-dependence (term $\cos(q_x/2)$) starts with order 7.
The extreme values of the cosine are at ${\bf{Q}}_1$ and ${\bf{Q}}_0$.
To quantify the variation of $S_{\bf{q}}$ we plot in Fig.~\ref{fig17}
the difference $\Delta S_{\mathbf{Q}}=S_{\mathbf{Q}_{1}}-S_{\mathbf{Q}_{0}}$ as a function of temperature.
We find indeed an {\it order by disorder} selection of the $\mathbf{Q}_{1}$ structure, 
although the magnitude of $\Delta S_{\mathbf{Q}}$ is small.
This result is in agreement with the findings of Canals and Lacroix \cite{Canals1998,Canals2000} 
and the corresponding spin structure is a collinear phase, 
where (classically) the total spin vanishes on each tetrahedron and neighboring tetrahedra are dephased by $\pi$.
We also find that $\Delta S_{\mathbf{Q}}$ is largest for the extreme quantum case. 
For larger spin quantum numbers $S=2$ and $3$ the curves $\Delta S_{\mathbf{Q}}/(S(S+1))$ versus $T/(S(S+1))$ almost coincide.
Let us mention here that the {\it order by disorder} selection due to thermal fluctuations discussed above 
is in accordance with the selection of collinear spin structures by quantum fluctuations 
found by large-$S$ approaches \cite{Henley_2006,Hizi_2007,Hizi_2009}, 
see also our discussion of the excitation spectrum in Sec.~\ref{sec4}.
	
\begin{widetext}
 
\begin{table}
\begin{centering}
\caption{First nine HTE terms of the static structure factor $s_{\mathbf{q},S,n}/((S(S+1))^{n+1}J^{n})$ for ${\mathbf{q}}=(q_x,4\pi,0)$
for the spin-$S$ PHAF with $S=1/2$, $S=1$, and $S=3/2$.
\label{tab1}}
\begin{tabular}{|c|c|c|c|}
\hline 
$n$ & $S=1/2$ & $S=1$ & $S=3/2$\tabularnewline
\hline 
\hline 
\multirow{2}{*}{$1$} & \multirow{2}{*}{$\frac{2}{3}$} & \multirow{2}{*}{$\frac{2}{3}$} & \multirow{2}{*}{$\frac{2}{3}$}\tabularnewline
 &  &  & \tabularnewline
\hline 
\multirow{2}{*}{$2$} & \multirow{2}{*}{$0$} & \multirow{2}{*}{$-\frac{5}{36}$} & \multirow{2}{*}{$-\frac{8}{45}$}\tabularnewline
 &  &  & \tabularnewline
\hline 
\multirow{2}{*}{$3$} & \multirow{2}{*}{$-\frac{20}{27}$} & \multirow{2}{*}{$-\frac{10}{27}$} & \multirow{2}{*}{$-\frac{2588}{10125}$}\tabularnewline
 &  &  & \tabularnewline
\hline 
\multirow{2}{*}{$4$} & \multirow{2}{*}{$\frac{62}{243}$} & \multirow{2}{*}{$\frac{1721}{5184}$} & \multirow{2}{*}{$\frac{8662}{30375}$}\tabularnewline
 &  &  & \tabularnewline
\hline 
\multirow{2}{*}{$5$} & \multirow{2}{*}{$\frac{1312}{1215}$} & \multirow{2}{*}{$\frac{133}{810}$} & \multirow{2}{*}{$\frac{84448}{3796875}$}\tabularnewline
 &  &  & \tabularnewline
\hline 
\multirow{2}{*}{$6$} & \multirow{2}{*}{$-\frac{28006}{32805}$} & \multirow{2}{*}{$-\frac{32309}{69120}$} & \multirow{2}{*}{$-\frac{142434998}{512578125}$}\tabularnewline
 &  &  & \tabularnewline
\hline 
\multirow{3}{*}{$7$} & \multirow{3}{*}{$\frac{-1031308-560\textrm{cos}\frac{q_{x}}{2}}{688905}$} & \multirow{3}{*}{$\frac{6039471-4480\textrm{cos}\frac{q_{x}}{2}}{39191040}$} & \multirow{3}{*}{$\frac{11132918004-1750000\textrm{cos}\frac{q_{x}}{2}}{53820703125}$}\tabularnewline
 &  &  & \tabularnewline
 &  &  & \tabularnewline
\hline 
\multirow{2}{*}{$8$} & \multirow{2}{*}{$\frac{4423862+3608\textrm{cos}\frac{q_{x}}{2}}{2066715}$} & \multirow{2}{*}{$\frac{1552120827+633088\textrm{cos}\frac{q_{x}}{2}}{3762339840}$} & \multirow{2}{*}{$\frac{552725758-498920\textrm{cos}\frac{q_{x}}{2}}{6458484375}$}\tabularnewline
 &  &  & \tabularnewline
\hline 
\multirow{2}{*}{$9$} & \multirow{2}{*}{$\frac{18947028+8576\textrm{cos}\frac{q_{x}}{2}}{11160261}$} & \multirow{2}{*}{$\frac{-573191935+107488\textrm{cos}\frac{q_{x}}{2}}{1128701952}$} & \multirow{2}{*}{$\frac{-31183199780044+38350832000\textrm{cos}\frac{q_{x}}{2}}{108986923828125}$}\tabularnewline
 &  &  & \tabularnewline
\hline 
\end{tabular}
\par\end{centering}
\end{table}

\end{widetext}

\section{Summary}
\label{sec6} 
\setcounter{equation}{0}

We have presented a comprehensive study of the ground-state and finite-temperature static and dynamical properties of the spin-$S$ PHAF
using a rotation-invariant Green's function method (RGM) and the high-temperature expansion (HTE). 
The focus of our study is on the extreme quantum cases $S=1/2$ and $S=1$.  

To summarize some of our findings, 
we mention first that within our approaches we do not find indications of magnetic long-range order for all temperatures $T\ge 0$, 
including the absence of ground-state magnetic long-range order for arbitrary $S$.  
Already at $T=0$ the spin-spin correlations are extremely short-ranged
leading to a correlation length that is below the nearest-neighbor separation.
It is appropriate to mention that by means of the PFFRG approach \cite{FPRG_Pyro_2018}
the analysis of the RG flow yields some indications for a finite-temperature transition for some intermediate values of $1 < S < \infty$.  
However, the authors of that study were finally unable to conclude 
about the presence (or absence) of magnetic long-range order and/or to determine the nature of the magnetic order (if any). 
In particular, 
in agreement with our study no divergence of the static structure factor $S_{\bf{q}}$ at any ${\bf{q}}$-vector was found.
Second, 
the RGM approach gives a temperature-dependent excitation spectrum. 
We find two degenerate flat-modes and two dispersive modes.
By contrast to the linear spin-wave theory \cite{Sobral1997} 
the flat modes are not the lowest ones, but approach zero energy as $S\to \infty$. 
Comparing our RGM energy dispersions at $T=0$ with linear spin-wave data of Ref.~\cite{Sobral1997}
one may conclude that the RGM data are in favor of collinear spin states.
Third, 
the static structure factor has ``spin-ice'' features seen as pinch points  \cite{Huang2016,FPRG_Pyro_2018} even at $T=0$.
Momentum cuts through the pinch points demonstrate that these points become sharper as $S$ increases.
Fourth,
the RGM data of the dynamical structure factor are applicable to interpret neutron scattering data 
for the $S=1$ pyrochlore compound NaCaNi$_2$F$_7$,
however, with the exception of the lowest frequencies.
Fifth, 
the HTE data for the $\mathbf{q}$-dependence of the static structure factor 
illustrate a weak {\it order by disorder} selection of a collinear spin structure
that emerges as the temperature goes down from the infinite-temperature limit.
The HTE analysis is rigorous within an appropriate (high) temperature range $ T/(S(S+1)) \lesssim J$   
and may be used further to detect favored magnetic structures due to small extra interactions.
Finally,
the reported temperature dependences of 
the spin correlations,
the specific heat, 
and 
the uniform susceptibility 
obtained by RGM and HTE 
may provide useful benchmarks for further study of these properties by other methods.

\section*{Acknowledgments}

We acknowledge useful discussions with Y.~Iqbal, P.~McClarty, and R.~Moessner.
J.~R. and O.~D. thank the Wilhelm und Else Heraeus Stiftung for the kind hospitality 
at the 673.~WE-Heraeus-Seminar ``Trends in Quantum Magnetism'' (Bad Honnef, 4-8 June 2018).
O.~D. acknowledges the kind hospitality of the MPIPKS, Dresden in April-June and September of 2018
and at the Workshop ``Correlated Electrons in Transition-Metal Compounds: New Challenges'' (5-9 November 2018).
The work of O.~D. was partially supported by Project FF-30F (No.~0116U001539) from the Ministry of Education and Science of Ukraine.

\section*{Appendix: First terms of the static structure factor within the HTE}
\label{secA}
\renewcommand{\theequation}{A.\arabic{equation}}
\setcounter{equation}{0}

In this appendix we provide explicit formulas for the first three terms of the HTE for the static structure factor.
For $S=1/2$ we have: 
\begin{align}
\label{app1}
\frac{S_{\mathbf{q},S=1/2}}{S(S+1)}= 1  
\nonumber\\
-\frac{2J}{3\tilde{T}}\left(\cos\frac{q_{x}}{4}\cos\frac{q_{y}}{4}+\cos\frac{q_{y}}{4}\cos\frac{q_{z}}{4}+\cos\frac{q_{x}}{4}\cos\frac{q_{z}}{4}\right)
\nonumber\\
+\frac{2J^{2}}{9\tilde{T}^{2}}\left(\cos\frac{q_{z}}{2}\cos\frac{q_{x}}{2}
                                                      +\cos\frac{q_{y}}{2}\cos\frac{q_{z}}{2}
                                                      +\cos\frac{q_{x}}{2}\cos\frac{q_{y}}{2}\right)
\nonumber\\
+\frac{2J^{2}}{9\tilde{T}^{2}}\cos\frac{q_{x}}{4}\left(2\cos\frac{q_{y}}{2}+1\right)\cos\frac{q_{z}}{4}
\nonumber\\
+\frac{2J^{2}}{9\tilde{T}^{2}}\cos\frac{q_{x}}{4}\cos\frac{q_{y}}{4}\left(2\cos\frac{q_{z}}{2}+1\right)
\nonumber\\
+\frac{2J^{2}}{9\tilde{T}^{2}}\left(2\cos\frac{q_{x}}{2}+1\right)\cos\frac{q_{y}}{4}\cos\frac{q_{z}}{4}
\nonumber\\
+\ldots+\frac{s_{\mathbf{q},S=1/2,12}}{(S(S+1))^{13}\tilde{T}^{12}}.
\end{align}
For $S=1$ we have:
\begin{align}
\label{app2}
\frac{S_{\mathbf{q},S=1}}{S(S+1)} = 1 
\nonumber\\
-\frac{2J}{3\tilde{T}}\left(\cos\frac{q_{x}}{4}\cos\frac{q_{y}}{4}+\cos\frac{q_{y}}{4}\cos\frac{q_{z}}{4}+\cos\frac{q_{x}}{4}\cos\frac{q_{z}}{4}\right)
\nonumber\\
+\frac{2J^{2}}{9\tilde{T}^{2}}\left(\cos\frac{q_{z}}{2}\cos\frac{q_{x}}{2}
                                                      +\cos\frac{q_{y}}{2}\cos\frac{q_{z}}{2}
                                                      +\cos\frac{q_{x}}{2}\cos\frac{q_{y}}{2}\right)
\nonumber\\
+\frac{2J^{2}}{9\tilde{T}^{2}}\cos\frac{q_{x}}{4}\left(2\cos\frac{q_{y}}{2}+\frac{13}{8}\right)\cos\frac{q_{z}}{4}
\nonumber\\
+\frac{2J^{2}}{9\tilde{T}^{2}}\cos\frac{q_{x}}{4}\cos\frac{q_{y}}{4}\left(2\cos\frac{q_{z}}{2}+\frac{13}{8}\right)
\nonumber\\
+\frac{2J^{2}}{9\tilde{T}^{2}}\left(2\cos\frac{q_{x}}{2}+\frac{13}{8}\right)\cos\frac{q_{y}}{4}\cos\frac{q_{z}}{4}
\nonumber\\
+\ldots+\frac{s_{\mathbf{q},S=1,10}}{(S(S+1))^{11}\tilde{T}^{10}}.
\end{align}
Finally, for $S=3/2$ we have:
\begin{align}
\label{app3}
\frac{S_{\mathbf{q},S=3/2}}{S(S+1)}= 1 
\nonumber\\
-\frac{2J}{3\tilde{T}}\left(\cos\frac{q_{x}}{4}\cos\frac{q_{y}}{4}+\cos\frac{q_{y}}{4}\cos\frac{q_{z}}{4}+\cos\frac{q_{x}}{4}\cos\frac{q_{z}}{4}\right)
\nonumber\\
+\frac{2J^{2}}{9\tilde{T}^{2}}\left(\cos\frac{q_{z}}{2}\cos\frac{q_{x}}{2}
                                                      +\cos\frac{q_{y}}{2}\cos\frac{q_{z}}{2}
                                                      +\cos\frac{q_{x}}{2}\cos\frac{q_{y}}{2}\right)
\nonumber\\
+\frac{2J^{2}}{9\tilde{T}^{2}}\cos\frac{q_{x}}{4}\cos\frac{q_{y}}{4}\left(2\cos\frac{q_{z}}{2}+\frac{9}{5}\right)
\nonumber\\
+\frac{2J^{2}}{9\tilde{T}^{2}}\cos\frac{q_{x}}{4}\left(2\cos\frac{q_{y}}{2}+\frac{9}{5}\right)\cos\frac{q_{z}}{4}
\nonumber\\
+\frac{2J^{2}}{9\tilde{T}^{2}}\left(2\cos\frac{q_{x}}{2}+\frac{9}{5}\right)\cos\frac{q_{y}}{4}\cos\frac{q_{z}}{4}
\nonumber\\
+\ldots+\frac{s_{\mathbf{q},S=3/2,10}}{(S(S+1))^{11}\tilde{T}^{10}}.
\end{align}
In the above equations the abbreviation $\tilde{T}=T/(S(S+1))$ is used.
The $S$-dependence of $S_{\mathbf{q}}$ appears first in terms of second order in $T$.
Setting $q_y=4\pi$ and $q_z=0$ in these formulas we reproduce the first rows from Table~\ref{tab1}.

\bibliography{phaf}

\end{document}